\documentclass[12 pt]{article}

\newcommand{\NaYb}{NaYbSe\textsubscript{2}}
\newcommand{\NaYbLu}{NaYb\textsubscript{$x$}Lu\textsubscript{$1-x$}Se\textsubscript{2}}
\newcommand{\NaLu}{NaLuSe\textsubscript{2}}
\newcommand{\KappaOrganic}{$\kappa$-(BEDT-TTF)\textsubscript{2}Cu\textsubscript{2}(CN)\textsubscript{3}}

\usepackage{CJK}

\usepackage[letterpaper, top = 1 in, bottom = 0.8 in, left = 0.8 in, right = 0.8 in]{geometry}

\usepackage{fancyhdr, lastpage}
\usepackage{framed}

\usepackage[explicit]{titlesec}

\titleformat{\section}%
{\normalfont}{\thesection}{1em}{\textbf{\MakeUppercase{#1}}}

\titleformat{\subsection}%
{\normalfont}{\thesubsection}{1em}{\textsc{#1}}

\usepackage{dashrule}
\newcommand{\Rule}{\noindent\rule{\linewidth}{0.5 pt}}
\newcommand{\DRule}{\noindent\hdashrule{\linewidth}{0.5 pt}{1pt}}

\usepackage{enumitem}

\setlist[description]{font = {\normalfont\bfseries}}

\usepackage[dvipsnames, svgnames, table]{xcolor}

\usepackage[]{hyperref}
\hypersetup{colorlinks = true, linkcolor = PineGreen, citecolor = blue, filecolor = magenta, urlcolor = blue}

\usepackage{graphicx}
\usepackage{float, caption, subcaption}

\newcommand{\FigRef}[1]{Fig.~\ref{#1}}

\usepackage{amsmath, amssymb, amsthm}
\usepackage[scr = boondoxo, scrscaled = 1.05]{mathalfa}
\numberwithin{equation}{section}

\allowdisplaybreaks

\usepackage{siunitx}
\usepackage{physics, braket}
\AtBeginDocument{\RenewCommandCopy\qty\SI}
\newcommand{\SIU}[2]{\SI[separate-uncertainty = true]{#1}{#2}}
\DeclareSIUnit{\angstrom}{\textup{\AA}}

\usepackage{tabularx, multirow}
\newcolumntype{Y}{>{\centering\arraybackslash}X}

\usepackage{tocloft}

\tocloftpagestyle{fancy}

\usepackage[numbers]{natbib}

\newcommand{\beginreferences}{%
	\phantomsection%
	\addcontentsline{toc}{section}{References}%
}

\newcommand{\beginsupplement}{
	\appendix
	
	\setcounter{section}{0}
	\renewcommand{\thesection}{S\arabic{section}}%
	\setcounter{table}{0}
	\renewcommand{\thetable}{S\arabic{table}}%
	\setcounter{figure}{0}
	\renewcommand{\thefigure}{S\arabic{figure}}%
	
	\phantomsection\section*{Supplementary Information}%
	
	\addcontentsline{toc}{section}{Supplementary Information}
	
}

\usepackage{authblk}
\date{\today}

\newcommand{\FirstAuthor}{L. Pritchard Cairns, Y. Lyu}

\newcommand{\Email}{analytis@berkeley.edu}
\newcommand{\AltEmail}{yuanqilyu@berkeley.edu}

\title{Entanglement Randomness and Gapped Itinerant Carriers in a Frustrated Quantum Magnet}

\author[*, 1]{\textbf{Luke Pritchard Cairns}}
\author[*, $\dagger$, 1]{\textbf{Yuanqi Lyu}}
\author[1, 2]{Josue Rodriguez}
\author[1]{Chunxiao Liu}
\author[1]{Kenneth Ng}
\author[3]{John Singleton}
\author[$\ddagger$, 1, 2, 4, 5]{James G. Analytis}

\affil[*]{Equal contribution.}
\affil[$\dagger$]{Email: \href{mailto:\AltEmail}{\textcolor{PineGreen}{\AltEmail}}}
\affil[$\ddagger$]{Corresponding author; Email: \href{mailto:\Email}{\textcolor{PineGreen}{\Email}}}

\affil[1]{\textit{Department of Physics, University of California, Berkeley, CA 94720, USA}}
\affil[2]{\textit{Materials Science Division, Lawrence Berkeley National Laboratory, Berkeley, CA 94720, USA}}
\affil[3]{\textit{National High Magnetic Field Laboratory, Los Alamos National Laboratory, Los Alamos, NM 87545, USA}}
\affil[4]{\textit{CIFAR Quantum Materials, Toronto, ON M5G 1M1, Canada}}
\affil[5]{\textit{Kavli Energy NanoScience Institute, Berkeley, CA 94720, USA}}

\makeatletter
\def\@maketitle{
	\begin{center}
		{\LARGE \textsc{\@title}}
		
		\vskip 1.5 em
		
		\@author{}

		\vskip 1.5 em
		
		\@date
		
	\end{center}
	\Rule{}
	\vskip 1.5 em
}
\makeatother

\begin{document}
	
	\maketitle
	\thispagestyle{fancy}
	
	\section*{Abstract}
	The quantum spin liquid (QSL) is a state manifesting extraordinary many-body entanglement, and the material \NaYb{} is thought to be one of the most promising candidates for its realization. Through low-temperature heat capacity and thermal conductivity measurements we identify an apparent contradiction familiar to many QSL candidates: while entropy is stored by apparently gapless excitations, the itinerant carriers of entropy are gapped. By studying the compositional series \NaYbLu{} across a percolation transition of the magnetic lattice, we suggest that this contradiction can be resolved by the presence of entanglement scales of random sizes. Moreover, as we truncate the scale of entanglement by magnetic dilution, we show that the itinerant magnetic entropy carrier in \NaYb{} is not the result of long-range entanglement but rather depends on the propagation of the simplest entangled object of all---the spin dimer.
	\clearpage
	
	\tableofcontents
	\clearpage
	
	\section{Main}
	Quantum entanglement is a many-body state that cannot be separated into the product of its single particle constituents. One simple example of an entangled spin ground state is a singlet dimer, formed when two $S=1/2$ spins interact through antiferromagnetic (AFM) Heisenberg exchange interaction. A quantum spin liquid (QSL) is a ground state of unparalleled many-body entanglement where all spins in a connected lattice are entangled together \cite{Broholm2020}. Among many interesting theoretical properties, such a system could manifest excitations known as spinons with anyonic statistics beyond the fermions and bosons characterizing the Standard Model.\cite{Savary2016} The realization of this state, however, is usually precluded by the onset of long range order, which is often favoured energetically \cite{Huse1988} and consequently the ground state in almost every case \cite{Zhu2015, Broholm2020}. In this case the wavefunction is effectively collapsed into a frozen configuration of spins that spontaneously break the underlying symmetry of the lattice. The many-body entangled state can be brought back into favour---at least in numerical simulations---when strong geometric frustration suppresses conventional ordering, thus allowing a QSL to form within a very specific parameter space of interaction strengths \cite{Zhu2015, Iaconis2018, Zhu2018}.
	
	The experimental observation of such an exotic state has proven challenging. Practically, the scale of entanglement is commonly limited by material disorder; lattice disorders and randomness not only release the geometric frustration locally, but also induce decoherence and collapse long-range entanglement via an ``ordering due to disorder'' mechanism \cite{Sheng1992}. In the past decade, it is shown both numerically \cite{Watanabe2014, Kawamura2019} and analytically \cite{Kimchi2018} that the interplay between entanglement and the randomness among the exchange interactions would stabilize---instead of a QSL---a valence bond glass (VBG) as the ground state. In lieu of long-range entanglement involving all spins as in a QSL, a VBG consists mostly of tiled spin-singlet dimers, as well as the occasional single orphan spins and locally-entangled larger-than-dimer clusters scattered among the dimer tilings.\cite{Kawamura2019} This state permits short-range order and entanglement to coexist over different length scales. The question of present interest is how the length scale of entanglement evolves between that of the humble dimer, the randomness of the VBG and ultimately one that could host a quantum spin liquid.
	
	Our subject is \NaYb{}, a prime candidate for QSL for its exemplary properties: Yb\textsuperscript{$3+$} ions of effective spin $1/2$ form 2-D equilateral triangular lattices where the $\mathscr{J}_1/\mathscr{J}_2$---the ratio between nearest neighbour (NN) and next NN exchange strengths---is calculated to favour a QSL ground state \cite{Dai2021, Scheie2024}. The experimental verification of a QSL in \NaYb{}, however, has yielded conflicting results, similar to the situation for most other QSL candidates \cite{Yamashita2009, Bordelon2019, Dai2021}. On one hand, the key signature of spinons in a gapless $U\left(1\right)$ QSL---unusual fermionic thermodynamics in an electrical insulator---appears evident in the heat capacity as a large Sommerfeld coefficient ($T$-linear behaviour) \cite{Ranjith2019}; and inelastic neutron scattering (INS) detects a continuum indicative of a spinon Fermi surface \cite{Dai2021}. On the other hand, this same excitation is absent in the thermal conductivity, and the INS data simultaneously exhibits peaks corresponding to short-range \SI{120}{\degree}-AFM order \cite{Dai2021}. The coexistence of short-range order and entanglement in INS suggests a mixing of various entanglement length scales, which is in contrast with the clear numerical predictions where a uniform phase---either long-range entanglement or AFM order---prevails as the ground state \cite{Zhu2018}; real-world spatial disorders and randomness therefore must play a significant role in determining the length scales of the entanglements and the corresponding spin ground state.
	
	In this study we conduct heat capacity and thermal conductivity measurements on the composition series of \NaYbLu{}, where non-magnetic Lu\textsuperscript{$3+$} ions are uniformly distributed with the Yb\textsuperscript{$3+$}, connecting through a magnetic percolation transition \cite{PritchardCairns2022}. While we cannot control the intrinsic randomness in the \NaYb{}, the introduction of Lu\textsuperscript{$3+$} ions adds tunable artificial disorder and dilutes the once-fully-connected triangular magnetic lattice. This allows us to impose spatial constraints on spin-spin entanglement and tip the intricate balance between randomness and entanglement---separating the excitations originating within regions of long-range entanglement from those produced by short-range interactions.
	
	With this approach, we are able to draw several key conclusions from our data. In heat capacity, a sharp peak around the temperature of $\sim\SI{2}{\kelvin}$ is observed throughout the composition sequence; the majority of entropy is released within, demonstrating the population dominance of a single type of two level system---the dimers. A much broader shoulder feature rises at lower temperatures for near-unitary $x$---in accordance with the enhanced formation of locally-entangled large clusters when magnetic dilution and spatial interruptions are minimal. These two features, along with their population evolutions, suggest VBG be the ground state. Meanwhile, a gapped magnetic entropy carrier can be identified in thermal conductivity: it emerges sharply around the percolation transition of the magnetic lattice and remains dominant at higher $x$---proving its itinerant nature and magnetic origin. Surprisingly, the population of these carriers peaks around the percolation transition, confirming that it cannot arise from any long-range entangled objects, but rather from short-range connectivity. We propose that this carrier arises from the non-spin carrying low-energy excitations of VBG, which ready emerges at the boundaries between different spin features. Given randomness and disorder are ubiquitous, our discovery could help explain similar behaviours and controversies common to many other QSL candidates.
	\clearpage
	
	\section{Results}
	\begin{figure}[t!]
		\begin{center}
			\includegraphics[width = 8.8 cm]{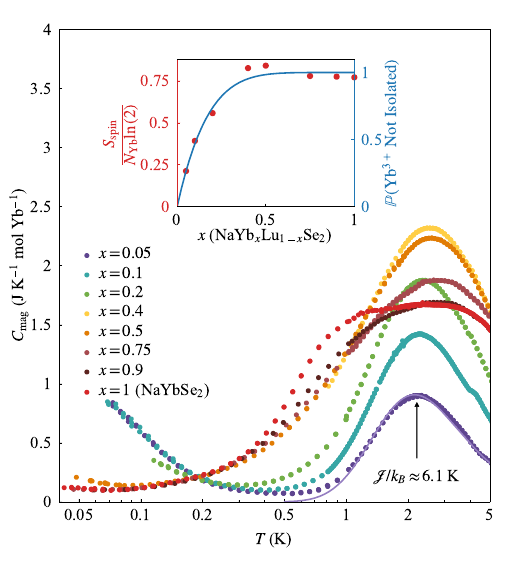}
		\end{center}
		
		\caption{%
			Zero-field magnetic heat capacity of \NaYbLu{}. The non-magnetic contribution has been subtracted, assuming it is identical to the \NaLu{} heat capacity in all cases \cite{PritchardCairns2022}. The solid line shows a fit to a Heisenberg model with $\mathscr{J}/k_B=6.1$~K for the $x=0.5$ compound, as described in the main text and SM. \textsc{Inset}: \textsc{Left Axis}: The integrated entropy release from electronic spins of the Yb\textsuperscript{$3+$} between 0.5 and \SI{30}{\kelvin}, plotted as a fraction of the anticipated entropy release according to the number of $J=1/2$ spins---$\ln(2)$ per spin. \textsc{Right Axis}: The probability that a Yb\textsuperscript{$3+$} ion will have at least one Yb\textsuperscript{$3+$} nearest neighbour---not isolated (see SI for details). It is plotted on a different scale compared to the left axis and not a fit of the data.
		}
		\label{Fig-HeatCapacity}
		\DRule
	\end{figure}
	
	Shown in \FigRef{Fig-HeatCapacity} is the zero-field magnetic heat capacity of the compositional series \NaYbLu{}. In the $x = 0.05$ compound the heat capacity peak is almost entirely attributable to isolated dimers collapsing into their respective singlet ground states. The position of the peak thus defines the energy scale of interaction, and the data is fit well by a Heisenberg model with $\mathscr{J}/k_B\approx\SI{6.1}{\kelvin}$. (Included in the supplementary material (SM) are more sophisticated attempts to constrain the spin Hamiltonian through modelling the heat capacity and magnetization of NaYb\textsubscript{0.05}Lu\textsubscript{0.95}Se\textsubscript{2}.)
	
	With increasing $x$---or the density of the magnetic Yb\textsuperscript{$3+$}---the probability for an isolated spin decreases. The heat capacity correspondingly releases more entropy, following the predicted population of connected spins (\FigRef{Fig-HeatCapacity} inset) and the expectation that isolated spins do not release entropy without an external field. The agreement between the measured entropy release and predicted population of connected spins further confirms that the mixture of Lu\textsuperscript{$3+$} and Yb\textsuperscript{$3+$} in our composition series are spatially uniform. By $x\geqslant 0.4$ almost all spins are connected to neighbouring magnetic sites and the released entropy saturates.
	
	In terms of the shape of the heat capacity curves, all measurements for $x\leqslant 0.5$ appear qualitatively similar to the dimer peak in NaYb\textsubscript{0.05}Lu\textsubscript{0.95}Se\textsubscript{2}, and all can be fit through a simple dimer model with continuously broadening energy levels with increasing Yb\textsuperscript{$3+$} ion density. However, it is important to note that a collection of small clusters (with appropriate concentrations according to $x$, see SM) can describe the data equally well. Regardless, at the opposite end of the series ($x=1$) the data appears qualitatively different, and a broad feature is observed at $\sim\SIU{1}{\kelvin}$, which turns into a shoulder feature with increased dilution ($x=0.75$ to $0.9$) before becoming unobservable below $x=0.5$. The position and breadth of this feature suggests that it originates from a magnetic structure with a broad energy spectrum, in contrast to the single energy level ($\mathscr{J}$) associated with the dimer singlet formation. Importantly, the heat capacity contribution from this broad hump is absorbed into the common feature as the system is diluted, as the length scale of connectivity decreases.  Given that the length scale of spin correlations is bounded by the continuity of the magnetic lattice---which is rapidly truncated by dilution--- this observation suggests the shoulder feature arises from clusters of short-range entangled spins.
	
	\begin{figure}[t!]
		\begin{center}
			\includegraphics[width = 12.1 cm]{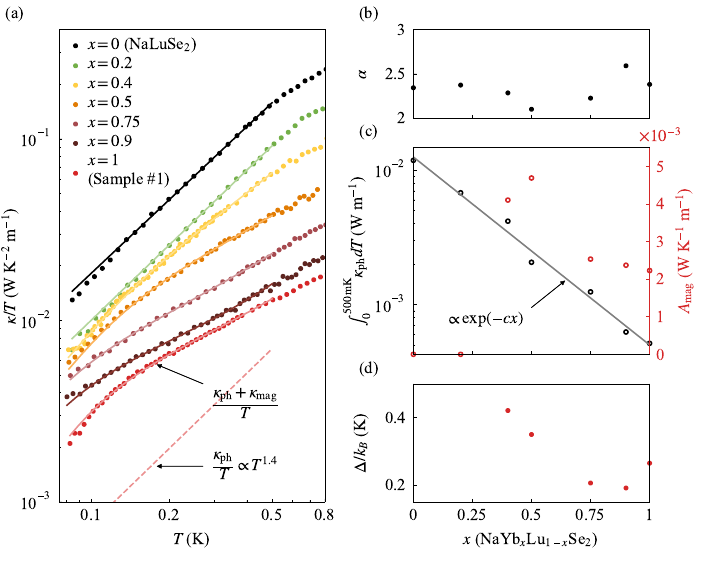}
		\end{center}
		
		\caption{%
			(a) Zero-field longitudinal thermal conductivity of \NaYbLu{}. The solid line is the fit with model described in the main text. The annotated red lines are fits of full thermal conductivity and phonon contributions of \NaYb{} respectively. (b), (c) and (d): The evolution of fit parameters. Here to avoid comparing $A_{\mathrm{ph}}$ of different units, we instead plot the integrated $\kappa_{\mathrm{ph}}$ from 0 to \SI{500}{\milli\kelvin}. The annotated grey line is an exponential fit of the integrated $\kappa_{\mathrm{ph}}$ with respect to $x$; $c$ is a fitting parameter.
		}
		\label{Fig-ThermalConductivity}
		\DRule
	\end{figure}
	
	Shown in \FigRef{Fig-ThermalConductivity} (a) are the zero-field longitudinal thermal conductivities across the compositional series. \NaLu{} matches the anticipated behaviour of a non-magnetic insulator, with a single $T^{2.3}$ power law below \SI{500}{\milli\kelvin}. This suppression from the anticipated $T^3$ phonon behaviour is consistent with the thermal conductivity measured in a variety of other non-magnetic insulators and compounds of similar structure \cite{Li2008, Kimchi2018, Ma2018}, and is discussed in the SM. The full magnetic compound \NaYb{} shows markedly different behavior in the same temperature range, with a thermal conductivity of a significantly smaller magnitude and a bump-like curvature at a low-temperature that does not conform to any single power law.
	
	\begin{figure}[t!]
		\begin{center}
			\includegraphics[width = 8.8 cm]{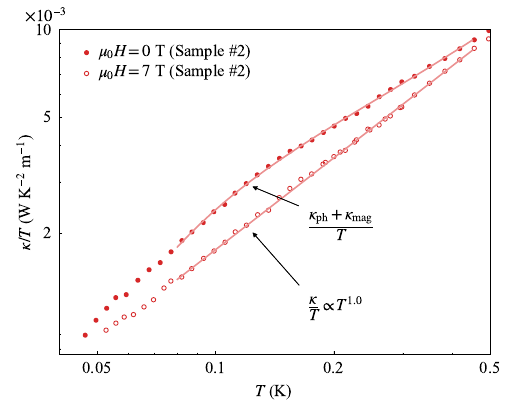}
		\end{center}
		
		\caption{%
			Thermal conductivity of \NaYb{} in zero-field and in an applied magnetic field of $\mu_0H=\SI{7}{\tesla}$ parallel to $c$-axis. The annotated lines are fits to thermal conductivity data, wherein we apply Eq.~\eqref{Eq-ThermalConductivity} for the zero-field data, and a temperature power law for the in-field data.
		}
		\label{Fig-InFieldData}
		\DRule
	\end{figure}
	
	When an external magnetic field of $\mu_0H =\SI{7}{\tesla}$ is applied parallel to the $c$-axis, as shown in \FigRef{Fig-InFieldData}, the thermal conductivity of \NaYb{} is suppressed and notably the bump-like curvature disappears, leaving behind a simple power law that can be attributed entirely to phonons. This implies that the bump-like curvature arises from some additional magnetic contribution that is suppressed by the application of the field. We thus model the zero-field thermal conductivity as a combination of a phonon power law $\kappa_{\mathrm{ph}}$ and a gapped magnetic itinerant entropy carrier $\kappa_{\mathrm{mag}}$, following reference \cite{Yamashita2009}:
	\begin{equation}\label{Eq-ThermalConductivity}
		\kappa=\kappa_{\mathrm{ph}}+\kappa_{\mathrm{mag}}=A_{\mathrm{ph}}T^\alpha+A_{\mathrm{mag}}\exp(-\frac{\Delta}{k_BT}) \text{.}
	\end{equation}
	We fit this model---with added consideration on spatial dimensionality of the magnetic carrier in the expression for  $\kappa_{\mathrm{mag}}$ (see \cite{SM} for details)---to the thermal conductivity of \NaYb{}. The gap is extracted to be $\Delta/k_B\approx\SI{270}{\milli\kelvin}$. The fitted phonon contribution has an exponent $\alpha$ that is similar both to that of \NaYb{} in-field and \NaLu{}, which further justifies our model. It is our expectation that the application of even larger fields would cause the thermal conductivity to rise as the spins become polarized and the phonon scattering is consequently reduced, as has been observed in previous studies on \NaYb{} \cite{Li2024} and related compounds \cite{Xu2016, Hong2023}.
	
	Applying Eq. \eqref{Eq-ThermalConductivity} to all thermal conductivity results, we obtain the fit parameters, shown in \FigRef{Fig-ThermalConductivity} (b), (c) and (d). For clarity, we describe separately the main trends observed in each fitting parameter:
	\begin{itemize}
		\item $A_{\mathrm{ph}}$ (\FigRef{Fig-ThermalConductivity} (c)) decreases exponentially with increasing density of magnetic Yb\textsuperscript{$3+$} sites. This suppression is dramatic; an order-of-magnitude decrease for all temperatures below \SI{1}{\kelvin}, showing very strong phonon scattering that could be linked to lattice disorder \cite{Klemens1958}. However, the exponential trend is quite peculiar---if it is the mixing of the Yb\textsuperscript{$3+$} and Lu\textsuperscript{$3+$} that generate such defects, the end members should be effectively less disordered than the mixed compounds. The systematic decrease with $x$ suggests the strong phonon scattering observed here must be intrinsic to the introduction of Yb\textsuperscript{$3+$} ions.
		\item $\alpha$ (\FigRef{Fig-ThermalConductivity} (b)) stays relatively constant within fitting uncertainty. This suggests that the phonon scattering evident in $A_{\mathrm{ph}}$ is relatively broadband and non-resonant with a specific energy scale (for instance, it cannot be due to interactions with the crystal field energy levels or a single exchange interaction strength). We will discuss the possible nature of such phonon scattering in the next section.
		\item $A_{\mathrm{mag}}$ (\FigRef{Fig-ThermalConductivity} (c))---amplitude of the magnetic contribution---is only non-zero for $x\geqslant 0.4$, a value that is very close to the percolation threshold of the triangular lattice (which theoretically occurs at $x = 0.5$), above which the Yb\textsuperscript{$3+$} lattice becomes connected. This sharp onset of $A_{\mathrm{mag}}$ around the percolation threshold hence proves that part of the spin excitation must be itinerant. Perhaps the most striking feature is that $A_{\mathrm{mag}}$ appears to peak near the percolation transition where the connected Yb\textsuperscript{$3+$} lattice is highly disordered: the corresponding magnetic carrier seems to prefer a disordered magnetic lattice over a pristine one.
		\item $\Delta$ (\FigRef{Fig-ThermalConductivity} (d)) shows a slight decreasing trend with increasing $x$. This suggests the origin of the gap is independent of the long-range lattice order, but rather linked to some local properties---at the scale of dimer formation.
	\end{itemize}
	\clearpage
	
	\section{Discussion}
	
	\begin{figure}[t!]
		\begin{center}
			\includegraphics[width = 8.8 cm]{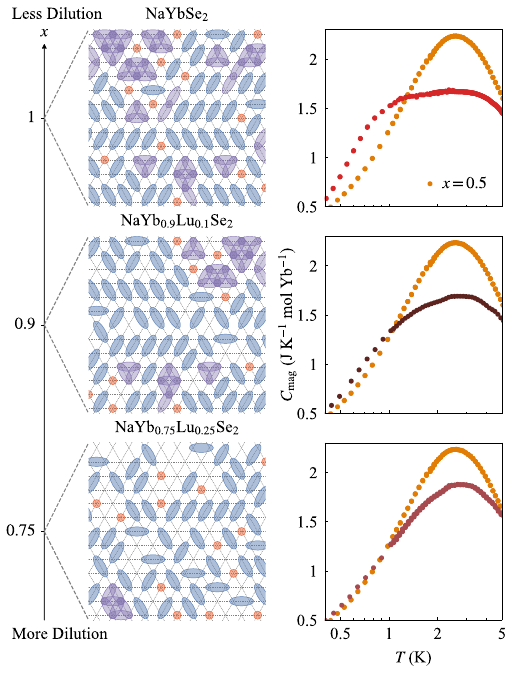}
		\end{center}
		\caption{%
			\textsc{Left}: Illustration of configurations of entangled spins at different compositions/magnetic site dilutions, where the red circles are orphan spins, blue ovals are spin-singlet dimers and purple regions are locally-entangled clusters. The unshaded nodes are Lu\textsuperscript{$3+$} ions, whose position is assigned randomly according to the expected composition. \textsc{Right}: The corresponding magnetic heat capacities of each composition. The heat capacity of $x=0.5$ (NaYb\textsubscript{0.5}Lu\textsubscript{0.5}Se\textsubscript{2}) is plotted as the orange points in each for comparison.
		}
		\label{Fig-CompositionEvolution}
		\DRule
	\end{figure}
	
\begin{figure}[t]
		\begin{center}
			\includegraphics[width = 18 cm]{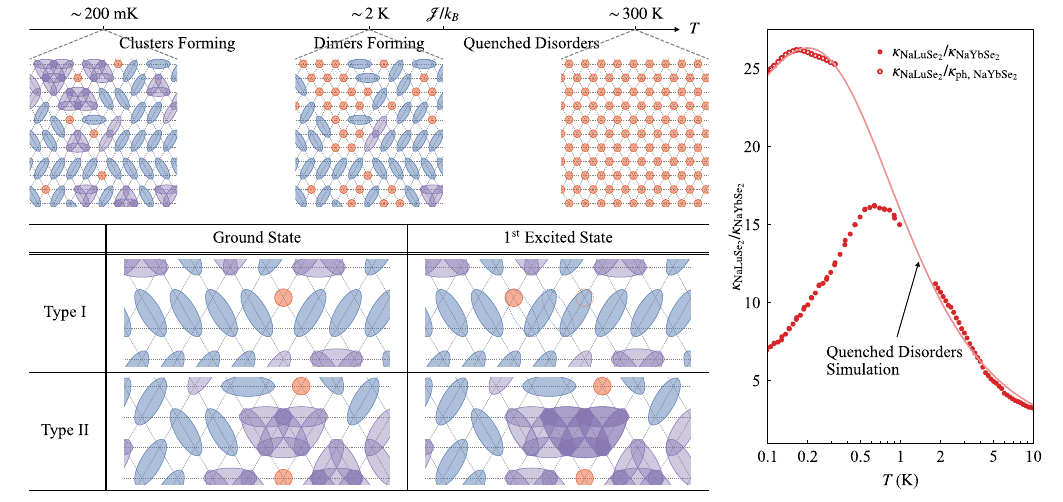}
		\end{center}
		
		\caption{%
			\textsc{Top}: Illustrations of the spin configurations in \NaYb{} at different temperatures, following the same color convention as in \FigRef{Fig-CompositionEvolution}. \textsc{Table}: Two types of low-energy excitations that can lead to an itinerant magnetic entropy carrier. In the right column, the original spin features are overlaid as dashed lines to visually assist the comparison. \textsc{Right}: The ratio of thermal conductivity between \NaLu{} and \NaYb{}, the open points are the ratio of thermal conductivity of \NaLu{} over the fitted phonon thermal conductivity $\kappa_{\mathrm{ph}}$ in \NaYb{}. The solid line is a simulation of the phonon thermal conductivity ratio given extra scattering due to quenched disorders in \NaYb{}.
		}
		\label{Fig-TemperatureEvolution}
		\DRule
	\end{figure}

	While \NaYb{} and \NaLu{} share almost identical structure, lattice constants \cite{PritchardCairns2022} and molecular weight, as plotted in \FigRef{Fig-TemperatureEvolution}, the phonon thermal conductivity $\kappa_{\mathrm{ph}}$ in \NaLu{} is $\sim2\times10^1$ times that of \NaYb{} below \SI{500}{\milli\kelvin}. Furthermore, the parameter $A_{\mathrm{ph}}$, which measures the phonon contribution to the thermal conductivity, decreases exponentially with increasing $x$. Such broadband scattering in temperature is unlikely to be caused by resonant scatterings between phonons and any magnetic transitions---consistent with the comparably small change in total thermal conductivity of \NaYb{} when an external field is applied (\FigRef{Fig-InFieldData}). Rather, the increased phonon scattering must arise intrisically from the interaction between individual Yb\textsuperscript{$3+$} ions and the lattice. Although the origin of this interaction is not known, it could arise from the strong spin-charge coupling typical in these materials that in turn leads to quenched disorder similar to other systems \cite{Watanabe2014,Abdel-Jawad2010, Abdel-Jawad2013}.  This strong phonon scattering is also observed in the related system YbMgGaO\textsubscript{4}, whose phonon thermal conductivity is about one fourth that of the non-magnetic LuMgGaO\textsubscript{4} \cite{Xu2016}---we include a discussion of their thermal conductivity in the SM. The important point for the present argument is that quenched randomness is likely intrinsic to the Yb\textsuperscript{$3+$} lattice, and decreases precipitously as the lattice is diluted of Yb\textsuperscript{$3+$} ions.
	
	Quenched lattice disorder translates into randomness in the exchange parameters. A number of previous studies \cite{Kimchi2018, Watanabe2014, Singh2010, Kawamura2019} have shown both numerically and analytically that the same (or very similar) random-bond-strength Hamiltonian will yield a Valence Bond Glass (VBG) ground state. This is characterized by a distribution of entangled objects of mostly spin-singlets tiled around a larger-than-dimer entangled clusters and orphan spins. The effect of dilution $x$ will only enhance the randomness of the exchange, stabilizing the VBG ground state \cite{Watanabe2014} and spin-glass physics in general \cite{Sheng1992, Liu2018, Ansari2024}.
	
	The coexistence of both dimers and clusters is vivid in the heat capacity data of \NaYb{}; which exhibits an initial peak at the same temperature that dimers form, followed by a broad shoulder feature at lower temperatures corresponding to the formation of larger clusters. As the addition of Lu\textsuperscript{$3+$} ions breaks up the fully-connected magnetic lattice, the length scale of spin-spin entanglement become physically bounded. Consequently, a sharp decrease in the population of the clusters---which are more space-sensitive---is observed as a rapid shrinking of the shoulder feature with decreasing $x$, leaving just an engorged dimer peak as the system is diluted. We illustrate the change in the ratios between dimers and clusters (akin to the VBG proposal) with different compositions in \FigRef{Fig-CompositionEvolution}, with the measured heat capacities plotted alongside. 
	
	The existence of an itinerant magnetic entropy carrier is evident in thermal conductivity from a comparison of the in-field and zero-field traces of \NaYb{}, as well as the emergence of the bump-like curvature near the percolation threshold. Intriguingly, the highest population of the itinerant magnetic carrier---the maximum of $A_{\mathrm{mag}}$ in \FigRef{Fig-ThermalConductivity} (c)---appears near the percolation transition, where the population of locally-entangled clusters is comparatively small. The itinerant magnetic entropy carrier is therefore unlikely to be a result of the entangled clusters, since these would be maximized at $x=1$, while the itinerant magnetic carriers are minimized---a direct illustration that the degrees of freedom that store heat differ from those that carry it. On the contrary, the itinerant magnetic carrier seems to prefer a connected but disordered magnetic lattice wherein entangled clusters are minimized and tiled dimers are the dominant magnetic feature. 
	
	This is consistent with the VBG picture whereby the majority of magnetic excitations (frozen dimers and clusters) are limited by their physical size and are not itinerant \cite{Kimchi2018}. However, this leaves open the question of the underlying nature of the magnetic carrier. Motivated by exact diagonalization studies of the low-energy excitations of the VBG in Ref. \cite{Kawamura2019}, we suggest an intuitive picture of ``entanglement retiling". There are two mechanisms we suggest are active. In the first case, as illustrated in \FigRef{Fig-TemperatureEvolution} table row I, a series of correlated dimers flip along an existing domain wall like dominoes; the flipping terminates at orphan spins \cite{Kawamura2019, Kimchi2018}. In the second case, as illustrated in \FigRef{Fig-TemperatureEvolution} table row II, a cluster is expanding by absorbing nearby dimers and orphan spins \cite{Kawamura2019}. Both processes allow entropy to travel without a spin-flip by shifting the boundary of correlated or entangled regions in a 1-dimensional fashion. This mechanism also explains the preference of this carrier to lattice disorder, since intuitively the disorder should encourage the creation of boundaries in the form of dimer tiling domains and cluster boundaries. A more detailed modeling of the thermal conductivity is included in \cite{SM}.

	Since these low-energy excitations do not necessarily require a spin-flip (singlet to triplet transition), their energy is not tied to the average nearest neighbour spin-spin exchange interaction strength $\mathscr{J}$ \cite{Kimchi2018}. Instead, the physical distinction between the ground state and the excited state is characterized by the bond configuration the dimers. The energy difference between the two ground and excited configurations is then set by the average exchange randomness between the neighbouring bonds \cite{Kimchi2018}. This is consistent with the trend of increasing gap size $\Delta$ in the thermal conductivity as one approaches the percolation transition and dilution deepens the random potential (\FigRef{Fig-ThermalConductivity} (c)). (See SI \cite{SM} for a more detailed discussion on the mechanism of the itinerant entropy carrier)
	
	When an external magnetic field is applied, the orphan spins and non-singlet ($J_{\mathrm{total}}\neq 0$) clusters become polarized. This lifting of degeneracies increases the energy differences among different spin configurations, discourages the mobility of individual magnetic features including boundaries and consequently reduces the population of itinerant low-energy magnetic entropy carriers. In thermal conductivity measurements, we see a drastic decrease of gapped itinerant magnetic excitations in applied fields and instead a clean power law attributable exclusively to phonons (\FigRef{Fig-InFieldData}). While at lower temperatures there might occur other magnetic transitions, understanding this is independent of our main conclusion. Entanglement re-tiling is sufficient to explain the most dramatic features seen in our data.
	
	The above picture suggests a resolution to the dilemma of how entangled objects that are localized due to disorder may nevertheless have an emergent, itinerant carrier arising from the motion of their boundaries in \NaYb{}. We establish this by recognizing an apparent contradiction that arises in many QSL candidates: a large low-temperature heat capacity that suggests a gapless excitation, but a thermal conductivity that appears gapped \cite{Yamashita2009, Xu2016, Ma2018, Ni2018, Rao2021, Hong2022, Hong2023, Hong2024, Tu2024}. In \NaYb{}, both of these aspects survive until the onset of a magnetic percolation transition, and an itineracy that is in fact enhanced by the presence of disorder with a gap that is significantly smaller than the exchange interaction. Indeed, in almost all QSLs facing similar challenges in reconciling the entropy storing and entropy carrying degrees of freedom, the gap sizes $\Delta$ are also smaller than the exchange interaction $\mathscr{J}$ (see SI \cite{SM} for further discussion and a comparison of related compounds). It seems likely therefore, that a general mechanism is at play, especially one that ultimately relies on the motion of boundaries defined by the dynamics of simplest entangled object of all, the spin-dimer.
	
	\section{Acknowledgement}
	We thank Joel E. Moore, Ehud Altman, Itamar Kimchi, Joseph Orenstein, Kamran Behnia and Vidya Madhavan for helpful discussions. This work was supported by the U.S. Department of Energy (DOE), Office of Science, Basic Energy Sciences, Materials Sciences and Engineering Division under contract DEAC02-05-CH11231 within the Quantum Materials program (KC2202). L.P.C., Y.L. and J.G.A. were supported by the EPiQS Initiative of the Gordon and Betty Moore Foundation through grant no. GBMF9067. C.L. acknowledges the fellowship support from the Gordon and Betty Moore Foundation through the Emergent Phenomena in Quantum Systems (EPiQS) program. Work at the National High Magnetic Field Laboratory was supported by NSF Cooperative Agreements No. DMR-1644779 and No. DMR-2128556, the DOE, and the State of Florida. J.S. acknowledges support from the DOE Basic Energy Sciences FWP ``Science of 100 T''.
	\clearpage
	
	\section{Methods}
	\begin{description}
		\item[Crystal Synthesis.] Detailed descriptions of the crystal growth, characterisation and evidence for homogeneity of magnetic and non-magnetic sites in the $0<x<1$ compounds can be found in \cite{PritchardCairns2022}. All measurements were performed on high-quality single crystals.
		
		\item[Heat Capacity Measurements.] Heat capacity measurements were performed using a Quantum Design\textsuperscript{\textregistered} PPMS Dynacool with \textsuperscript{3}He insert option for temperatures $T\gtrsim\SI{1}{\kelvin}$ and a home-built setup in Bluefors\textsuperscript{\textregistered} LD250 dilution refrigerator for $T\lesssim\SI{1}{\kelvin}$. The data was acquired using the relaxation time method \cite{Bachmann1972} across the full temperature range, and analysed in the low-temperature region using a modified version of the full temperature response analysis described in \cite{Hwang1997}.
		
		\item[Thermal conductivity.] Thermal conductivity measurements were performed using a home-built apparatus in Bluefors\textsuperscript{\textregistered} LD250 dilution refrigerator for temperatures $T\lesssim\SI{1}{\kelvin}$, and in Quantum Design\textsuperscript{\textregistered} PPMS Dynacool for $T\gtrsim\SI{2}{\kelvin} $. A standard steady-state one heater, two thermometers method is employed in all measurements.
	\end{description}

	\clearpage
	\beginsupplement
	
	\section{Phonon Scattering}
	\subsection{Scattering from Sample Boundary in \texorpdfstring{\protect\NaLu{}}{NaLuSe2}}
	
	The thermal conductivity contributed by phonons can be modeled by---in the simplest form---the kinetic formula, as the product of the phonon's volumetric heat capacity $C_{\mathrm{ph}}$, mean free path $l_{\mathrm{ph}}$ and velocity---the speed of sound $v_s$, i.e.:
	\begin{equation}\label{Eq-KineticFormula}
		\kappa_{\mathrm{ph}}\left(T\right)=\frac{1}{3}C_{\mathrm{ph}}\left(T\right)l_{\mathrm{ph}}\left(T\right)v_s \text{.}
	\end{equation}
	The phonon's volumetric heat capacity can be calculated using the Debye model:
	\begin{equation}
		C_{\mathrm{ph}}\left(T\right)=9Nk_B\left(\frac{T}{\Theta}\right)^3\int_{x=0}^{\Theta/T}\frac{x^4\exp(x)}{\left[\exp(x)-1\right]^2}dx \text{,}
	\end{equation}
	wherein $x$ is the dimensionless phonon frequency---defined from the actual frequency $\omega$ as:
	\begin{equation}
		x=\frac{\hbar\omega}{k_BT} \text{,}
	\end{equation}
	$N$ is the volumetric density of the atoms, and $\Theta$ is the Debye temperature originating from the linearisation of the phonon dispersion---it is related to crystal properties as \cite{Kittel2005}:
	\begin{equation}
		\Theta=\frac{\hbar v_s}{k_B}\sqrt[3]{6\pi^2N} \text{.}
	\end{equation}
	In \NaLu{}, a fitting of the heat capacity yields a Debye temperature of about \SI{259}{\kelvin}; then given the unit cell size---measured by powdered powder X-ray diffraction (PXRD) as \SI{98.7}{\cubic\angstrom}---we estimate the speed of sound in \NaLu{} to be about \SI{2.5e3}{\meter\per\second} \cite{PritchardCairns2022}.
	
\begin{figure}[t!]
		\begin{center}
			\includegraphics[width = 8.8 cm]{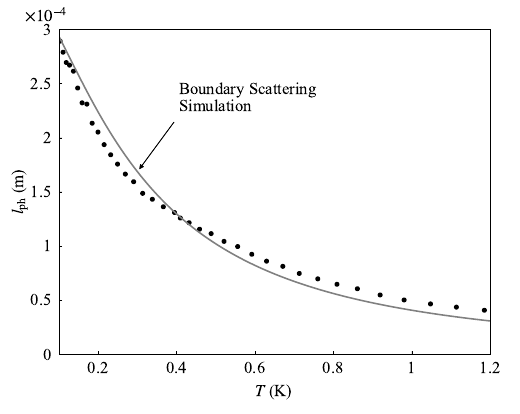}
			\caption{%
				The effective phonon mean free path of \NaLu{} calculated from the kinetic formula Eq.\eqref{Eq-KineticFormula}. The solid line is a simulation given the effect of specular boundary scattering.
			}
			\label{Fig-NaLuSe2PhononMeanFreePath}
		\end{center}
		\DRule
	\end{figure}

	At temperatures much lower than the Debye temperature---when $\Theta/T$ is effectively infinite for the integral---a $T^3$-dependency should emerge in the phonon heat capacity. This model fits well to the heat capacity of \NaLu{} below $\sim\SIU{40}{\kelvin}$. Combined with the estimated speed of sound, the fitted heat capacity allows us to obtain the effective phonon mean free path using the kinetic formula, as shown in \FigRef{Fig-NaLuSe2PhononMeanFreePath}. A strong temperature dependency can be seen from the plot, in accordance with the thermal conductivity not following a $T^3$-temperature power law.
	
	Up until now, we have assumed that the phonon mean free path has no dependence on frequency or polarization---thus we can pull it out of the heat capacity integral. This simplistic assumption breaks down, however, once we take the scattering of phonons into account. 
	
	For example, the lattice anharmonicity---which often appears as a non-linearity in the crystal elasticity---allows phonons near the boundaries of the first Brillouin zone to scatter among themselves via the Umklapp process. This leads to a decrease in the phonon mean free path at high temperature as the phonon density and mean frequency increase. The effect of the Umklapp process can be modeled as an reduction of the mean free path with increasing temperature and frequency, as \cite{Klemens1958}:
	\begin{equation}
		\frac{1}{l_{\text{ph, U}}\left(T,\omega\right)}\sim T\omega^2\sim T^3x^2 \text{.}
	\end{equation}
	It needs to be emphasized that this approximation of the Umklapp process only works for temperatures on par with or higher than the Debye temperature; the effect of Umklapp scattering is exponentially small for lower temperatures \cite{Klemens1958}.
	
	To estimate the thermal conductivity we shall now introduce the mean free path into the frequency integral, which gives the Callaway model \cite{Callaway1959}:
	\begin{equation}\label{Eq-CallawayModel}
		\kappa_{\mathrm{ph}}\left(T\right)=3Nk_Bv_s\left(\frac{T}{\Theta}\right)^3\int_{x=0}^{\Theta/T}\frac{x^4\exp(x)}{\left[\exp(x)-1\right]^2}l_{\mathrm{ph}}\left(T,x\right)dx\text{.}
	\end{equation}
	Since the mean free path is governed by the average time between phonon scattering events---i.e., the relaxation time---for a system with multiple scattering sources, Matthiessen's rule should be applicable, i.e.:
	\begin{equation}\label{Eq-MatthiessenRule}
		\frac{1}{l_{\mathrm{ph}}\left(T,x\right)}=\sum_{\text{All Sources}} \frac{1}{l_{\text{ph, Source}}\left(T,x\right)} \text{.}
	\end{equation}
	From the Callaway model it is clear that the Umklapp process nullifies the $T^3$-term at high temperature, leaving behind a $1/T$-temperature dependency, largely controlled by the upper bound of the definite integral.
	
	For insulators with a low defect density, phonon scattering within the bulk of the crystal should be minuscule at low temperatures, and consequently the mean free path could be close to or even exceed the physical dimensions of the sample. For our \NaLu{} sample, the width and thickness are about \SI{500}{\micro\meter} and \SI{20}{\micro\meter} respectively---both smaller than the effective mean free path below $\sim\SI{1}{\kelvin}$. As a result, the dominant phonon scattering source in \NaLu{} should be the crystal boundaries.
	The crystal boundary scattering, however, is a quite complicated subject to understand. Here we propose a phenomenological model that accounts for both diffusive and specular boundary scatterings: 
	\begin{itemize}
		\item On the one hand, if the sample surface is rough for most phonons---which implies a large roughness spacing as compared to the dominant phonon wavelength---we should expect the phonon scattering to be diffusive, and the phonons to be absorbed once they reach the boundaries. The resulting mean free path should therefore be some constant multiple of the sample dimensions, independent of either phonon frequencies or temperatures. Correspondingly, the phonon thermal conductivity should follow the $T^3$-temperature power law dictated by the phonon heat capacity. 
		\item On the other hand, if the surface is relatively smooth---with a roughness spacing comparable or smaller than the mean wavelength of the phonons---specular scattering then would selectively allow long-wavelength phonons to mirror reflect from the crystal boundaries, thus effectively extending the phonon mean free path far beyond the sample dimensions. Further, surface irregularities of a smaller size would have a stronger frequency-dependency when it comes to phonon scattering: point-like defects could contribute to right-hand side of Eq.~\eqref{Eq-MatthiessenRule} as $\omega^4$, shallow scratches as $\omega^3$, and growth steps as $\omega^2$ \cite{Klemens1958}. As a result, the phonon thermal conductivity is enhanced at low temperatures---as the dominant phonon frequencies decrease---and could behave as an effective temperature power law that deviates from $T^3$. 
	\end{itemize}
	Both scenarios have been observed in reference \cite{Li2008}, wherein the authors observe this deviation from $T^3$ in a good quality crystal, but then roughen the surface of the same sample and see a large decrease in the phonon thermal conductivity but a recovery of the $T^3$-temperature dependency.
	
	A quantitative modeling of the phonon boundary scattering can be derived as such. First, we introduce a constant $l_0^{-1}$ in the Matthiessen's summation to account for some strength of diffusive scattering. Next, as our sample surface is relatively smooth and our measurements are conducted at very low temperature, the mean phonon frequency is low. We therefore add a $\omega^2$---or equivalently $T^2x^2$---term to account for specular scattering, and the mean free path becomes:
	\begin{equation}
		\frac{1}{l_{\mathrm{ph}}\left(T,x\right)}\approx\frac{1}{l_{\text{ph, B}}\left(T,x\right)}=l_0^{-1}+bT^2x^2 \text{.}
	\end{equation}
	The effective phonon mean free path calculated through this model is overlaid on top of data in \FigRef{Fig-NaLuSe2PhononMeanFreePath}, where $l_0$ and $b$ are free parameters. Here we did not fit data to these parameters, but simply just picked values that gives a qualitative match.
	
	\subsection{Scattering from Quenched Randomness in \texorpdfstring{\NaYb{}}{NaYbSe2}}
	Now let us switch our focus onto \NaYb{}. While we cannot characterize the quenched random disorders in \NaYb{} directly, we can model its effect in terms of phonon scattering in the same phenomenological fashion that we treat boundary scatterings.
	
	More specifically, the quenched lattice distortions surrounding each Yb\textsuperscript{$3+$} ion generates a random local strain field, which induces a local change in elasticity and thus deflects phonons. Given this change is localized around each Yb\textsuperscript{$3+$} ion, the deflection effect should be stronger on phonons whose wavelengths are shorter. These characteristics of a quenched distortion are the same as the strain fields introduced by lattice dislocations \cite{Klemens1958}, which allows us to model its effect on phonon mean free path as:
	\begin{equation}
		\frac{1}{l_{\text{ph, D}}\left(T,x\right)}=l_1^{-1}+aTx \text{.}
	\end{equation}
	Here the $Tx$ term provides the wavelength dependency, while the $l_1^{-1}$ term accounts for some added diffusive scattering from the quenched disorder---similar to the analysis for boundary scattering.
	
	Consequently, the ratio of phonon thermal conductivities between \NaLu{} and \NaYb{} can be modeled as:
	\begin{equation}
		\begin{split}
			\frac{\kappa_{\text{\NaLu{}}}}{\kappa_{\text{ph, \NaYb{}}}}&=\frac{\displaystyle{3Nk_Bv_s\left(\frac{T}{\Theta}\right)^3\int_{x=0}^{\Theta/T}\frac{x^4\exp(x)}{\left[\exp(x)-1\right]^2}l_{\mathrm{ph, B}}\left(T,x\right)dx}}{\displaystyle{3Nk_Bv_s\left(\frac{T}{\Theta}\right)^3\int_{x=0}^{\Theta/T}\frac{x^4\exp(x)}{\left[\exp(x)-1\right]^2}\left[\frac{1}{l_{\mathrm{ph, B}}\left(T,x\right)}+\frac{1}{l_{\mathrm{ph, D}}\left(T,x\right)}\right]^{-1}dx}} \\
			&=\frac{\displaystyle{\int_{x=0}^{\Theta/T}\frac{x^4\exp(x)}{\left[\exp(x)-1\right]^2}\left(l_0^{-1}+bT^2x^2\right)^{-1}dx}}{\displaystyle{\int_{x=0}^{\Theta/T}\frac{x^4\exp(x)}{\left[\exp(x)-1\right]^2}\left[\left(l_0^{-1}+bT^2x^2\right)+\left(l_1^{-1}+aTx\right)\right]^{-1}dx}}\\
			&\approx\frac{\left(l_0^{-1}+l_1^{-1}\right)+AT+BT^2}{l_0^{-1}+BT^2} \text{;}
		\end{split}
	\end{equation}
	where the approximation in the last line holds when temperature is low and $\Theta/T$ goes to infinity. $A$ and $B$ are constants which result from the evaluation of the definite integrals---multiples of $a$ and $b$ respectively. A plot of this equation (without approximation) is shown as a solid line in the right panel of \FigRef{Fig-TemperatureEvolution}---with $l_1$ and $a$ picked to reflect the data; the values of $l_0$ and $b$ are the same as those used in \FigRef{Fig-NaLuSe2PhononMeanFreePath}.
	
	The approximate form of this ratio gives a rather clear picture of the temperature dependency of the suppression in the phonon thermal conductivity:
	\begin{itemize}
		\item For low temperature, the ratio approaches $\left(1+l_0\left/l_1\right.\right)$---a temperature-independent constant. This enables the $\kappa_{\mathrm{ph}}$ across the \NaYbLu{} composition series to follow similar effective temperature power laws below \SI{500}{\milli\kelvin}, as demonstrated in \FigRef{Fig-ThermalConductivity} (b). The overall suppression at low temperatures can then be attributed to the enhanced diffusive scattering from the random strain field.
		\item For intermediate temperatures, the ratio is dominated by the $T$-term in the numerator. As a result, it first rises slightly on top of the aforementioned constant, then drops akin to $1/T$. This explains the sharp-then-steady decrease in phonon thermal conductivity suppression as the temperature rises near and above \SI{1}{\kelvin}. 
		\item For high temperatures the $T^2$-terms dominate both in the numerator and the denominator, which leads to the ratio asymptotically approaching unitary.
	\end{itemize}	
	
	Since the induced local strain field is spatially random, with higher density of Yb\textsuperscript{$3+$}---or larger $x$---the probability of phonon scattering due to disorder get enhanced linearly with $x$. The phonon thermal conductivity consequently decreases in a Beer-Lambert fashion following an exponential curve with respect to $x$, as illustrated in \FigRef{Fig-ThermalConductivity} (c). The phonon thermal conductivity in \NaYb{}---as the right end of the exponential curve---gets suppressed so much that it falls inside the so-called ``glassy range'' \cite{Pohl2002}---wherein thermal conductivities of most amorphous solids reside and display temperature power laws between $T^2$ and $T^3$. 
	
	While it might be jarring at first sight that phonon thermal conductivity in \NaLu{} is over $\sim2\times10^1$ times higher than that in \NaYb{}, heightened phonon scattering due to induced glassy lattice randomness has been measured in a slew of salt mixtures wherein glassiness is engineered through tunable compositions \cite{Cahill1992, Pohl2002}. In (NaCl)\textsubscript{$1-x$}(NaCN)\textsubscript{$x$}, for example, the ratio between the thermal conductivity of $x=0$---pure NaCl---over that of $x=0.76$ peaks above $10^2$ and remains relatively constant from the lowest measured temperature around \SI{100}{\milli\kelvin} up to \SI{10}{\kelvin}; the ratio falls with higher temperatures and drops below 10 when the temperature rises above \SI{100}{\kelvin} \cite{Cahill1992}---matching the predictions from our model neatly.
	
	\begin{figure}[t!]
		\begin{center}
			\includegraphics[width = 12.1 cm]{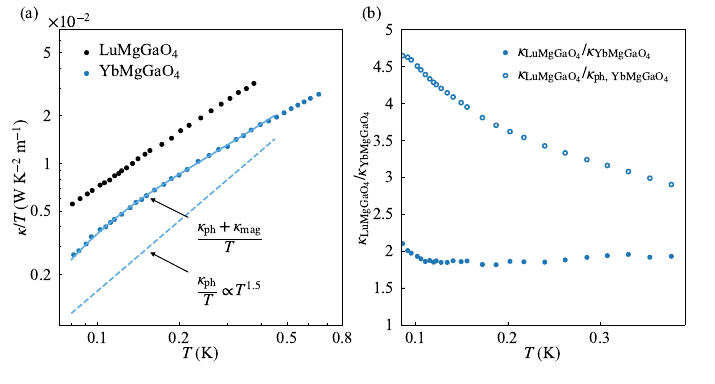}
			\caption{%
				(a) Thermal conductivities of LuMgGaO\textsubscript{4} and YbMgGaO\textsubscript{4}, data extracted from \cite{Xu2016}. Line plots are the fit results using the same model applied to \NaYbLu{}. (b) The ratio of thermal conductivity between LuMgGaO\textsubscript{4} and YbMgGaO\textsubscript{4}, the open points are the ratio of thermal conductivity of LuMgGaO\textsubscript{4} over the fitted phonon thermal conductivity $\kappa_{\mathrm{ph}}$ in YbMgGaO\textsubscript{4}.
			}
			\label{Fig-YbMgGaO4ThermalConductivity}
		\end{center}
		\DRule
	\end{figure}
	
	Similar phonon thermal conductivity suppression has also been observed in the pair LuMgGaO\textsubscript{4} and YbMgGaO\textsubscript{4}, with near identical lattice structures. YbMgGaO\textsubscript{4} is a QSL candidate wherein Yb\textsuperscript{$3+$} ions of effective $J=1/2$ form a 2-D triangular antiferromagnetic spin lattice, similar to \NaYb{} \cite{Xu2016, Li2017, Ma2018}. We extracted the thermal conductivity data from reference \cite{Xu2016} and applied the same fitting models used for \NaYbLu{}. The ratio of phonon thermal conductivity in LuMgGaO\textsubscript{4} over that in YbMgGaO\textsubscript{4} is about 4, as shown in \FigRef{Fig-YbMgGaO4ThermalConductivity}---we cannot fit the model given limited data availability. It needs to be emphasized that there is intrinsic lattice randomness in both LuMgGaO\textsubscript{4} and YbMgGaO\textsubscript{4} in the form of random site mixing between Mg\textsuperscript{$2+$} and Ga\textsuperscript{$3+$} \cite{Li2017}. The quenched randomness due to Yb\textsuperscript{$3+$} ions hence needs to ``compete'' with these intrinsic disorders, dulling its phonon thermal conductivity suppression effect. As a mater of fact, thermal conductivities of both LuMgGaO\textsubscript{4} and YbMgGaO\textsubscript{4} fall within the ``glassy range'', whereas for the \NaYb{}/\NaLu{} pair, only that of \NaYb{} is in the ``glassy range''.
	
	\clearpage
	
	\section{Spin-Charge Interactions \texorpdfstring{\\}{} and a Possible Origin of the Quenched Randomness}
	The lattice randomness in \NaYb{} and its dependency on $x$ across \NaYbLu{}---measured by phonon thermal conductivity---alludes to Yb\textsuperscript{$3+$} ions being the culprit. Hence we provide a brief overview of the spin-charge/lattice interactions in \NaYb{}.
	
	The energy-scale landscape of \NaYb{} has been well studied \cite{Zhang2021-Model, Scheie2024-KandNaYbSe2, Zangeneh2019, Schmidt2021}: The strong spin-orbit coupling first splits the energy levels of Yb\textsuperscript{$3+$} ($4f^{13}$) into $^2F_{5/2}$ and $^2F_{7/2}$ manifolds, with an energy gap of $\sim\SI{1e4}{\kelvin}k_B$ \cite{Zangeneh2019}. The $^2F_{7/2}$ manifold is then further split into 4 pairs of Kramer's doublets through strong spin-charge interaction in the form of crystal electric field (CEF) \cite{Zhang2021-CEF, Schmidt2021}. The doublet of the lowest energy is separated from other CEF levels by a gap of $\sim\SI{180}{\kelvin}k_B$---measured via inelastic neutron scattering (INS)---and thus effectively behaves as spin-$1/2$ for low temperatures and applied magnetic fields \cite{Zhang2021-CEF, Schmidt2021}.
	
	If we assume that each Yb\textsuperscript{$3+$} ion resides at centre of a perfect octahedron constituting of nearest neighbor Se\textsuperscript{$2-$} ions, the CEF Hamiltonian should follow cubic symmetry with only 2 free parameters \cite{Schmidt2021}:
	\begin{equation}
		\hat{\mathscr{H}}_{\text{CEF, Cubic}}=B^3_4\left(\hat{O}^0_4-20\sqrt{2}\hat{O}^3_4\right)+B^3_6\left(\hat{O}^0_6+\frac{35\sqrt{2}}{4}\hat{O}^3_6+\frac{77}{8}\hat{O}^6_6\right) \text{;}
	\end{equation}
	where $\hat{O}_\ell^m$ are the Steven's operators (see \cite{Hutchings1964} for explicit expressions). In reality, the Se\textsuperscript{$2-$}-octahedrons are distorted and a model with trigonal $C_\text{3v}$ symmetry is used in literatures to fit the CEF levels measured by INS \cite{Zhang2021-Model, Schmidt2021, Scheie2024}:
	\begin{equation}\label{Eq-CEFTrigonal}
		\hat{\mathscr{H}}_{\text{CEF}}=B^0_2\hat{O}^0_2+B^0_4\hat{O}^0_4+B^3_4\hat{O}^3_4+B^0_6\hat{O}^0_6+B^3_6\hat{O}^3_6+B^6_6\hat{O}^6_6\text{.}
	\end{equation}
	The additional $B^0_2\hat{O}^0_2$ allows distortions along one of the trigonal axes of the octahedron.
	
	In the case of \NaYb{}, we argue such distortions are not spatially homogeneous, and thus generates randomness in the lattice. Besides the exceptionally low ``glassy range'' phonon thermal conductivity, the lattice randomness in \NaYb{} is evident given the heightened widths of the CEF levels measured through INS at low temperatures. The widths are comparable to those observed in YbMgGaO\textsubscript{4} \cite{Dai2021, Li2017}---wherein the randomness in CEF levels has been well recognized and attributed to lattice disorders \cite{Li2017}. Similarly broad widths of the CEF levels are observed in KYbSe\textsubscript{2}, wherein the widths are qualitatively unchanged below \SI{100}{\kelvin} and the remain broad even at \SI{7}{\kelvin}---a temperature that is well below the equivalent temperatures of any CEF levels \cite{Scheie2024}.
	
	Similar quenched randomness has been reported in both \KappaOrganic{} and EtMe\textsubscript{3}-Sb[Pd(dmit)\textsubscript{2}]\textsubscript{2} \cite{Watanabe2014}, where it manifests as a glassy response in dielectric constants at temperatures much higher than that of spin freezing \cite{Abdel-Jawad2010, Abdel-Jawad2013}: it seems like in all these QSL candidates with geometrically frustrated lattices, the charge degree of freedom develops randomness at temperatures much higher than that corresponding to spin-spin interaction strength \cite{Watanabe2014}. Such universality hints at spin-charge coupling being the origin of the randomness. Indeed, the strong spin-charge coupling allows the CEF splittings to couple to lattice strengths, for instance \cite{Mullen1974}:
	\begin{equation}
		\begin{split}
			\hat{\mathscr{H}}'_{c_{11}-c_{12}}&=-g_2\sqrt{c_{11}^0-c_{12}^0}\left(\frac{2\epsilon_{zz}-\epsilon_{xx}-\epsilon_{yy}}{\sqrt{6}}\hat{O}_0^2+\frac{\epsilon_{xx}-\epsilon_{yy}}{\sqrt{2}}\hat{O}_2^2\right) \\
			&=-g_2\sqrt{c_{11}^0-c_{12}^0} \\
			&\qquad\times\left[\frac{2\epsilon_{zz}-\epsilon_{xx}-\epsilon_{yy}}{\sqrt{6}}\left(\sqrt{3}\hat{J}_x^2-\sqrt{3}\hat{J}_y^2\right)+\frac{\epsilon_{xx}-\epsilon_{yy}}{\sqrt{2}}\left(2\hat{J}_z^2-\hat{J}_x^2-\hat{J}_y^2\right)\right] \text{;}
		\end{split}
	\end{equation}
	where $\epsilon_{ij}$ are the lattice strain tensors, $c_{ij}^0$ are the elasticity constants and $g_2$ is the coupling strength.
	
	In other words, it is energetically favourable to shift the CEF levels of individual Yb\textsuperscript{$3+$} ions and generate local lattice strains and the corresponding distortions along any of the symmetry axes of each Se\textsuperscript{$2-$} octahedron. Since the distortions only relies on local interactions at the sites of individual Yb\textsuperscript{$3+$} ions, given the underlying geometric frustration---which forces neighbouring lattice stresses to bump into each other---its spatial distribution should be randomized and glassy. At the temperatures of spin freezing, these lattice distortions should be frozen in place as the quenched randomness. Such mechanism---as our educated guess for the origin of the quenched disorder---could help explain the unusually broad INS spectra and the strong phonon thermal conductivity suppressions.
	
	\clearpage
	
	\section{Itinerant Magnetic Entropy Carriers}
	\subsection{The Spatial Dimensionality}
	First, let us discuss the dimensionality of the itinerant magnetic entropy carrier. From the discussion in the main text---as well as the table in \FigRef{Fig-TemperatureEvolution}---the carrier should travel in 1-dimensional space. We can confirm this by comparing the modelling of thermal conductivity in different dimension spaces with our experimental data.
	
	More specifically, if we assume that the dispersion relation $\epsilon\left({\mathbf k}\right)$ of the carrier can be approximated linearly---akin to that of phonon or AFM magnon \cite{Kittel2005}---the density of states ${\mathscr D}\left(\epsilon\right)$ should follow:
	\begin{equation}
		{\mathscr D}\left(\epsilon\right)=\frac{d{\mathscr N}\left(\epsilon\right)}{d\epsilon}\sim\frac{d\epsilon^D}{d\epsilon}\sim\epsilon^{D-1} \text{;}
	\end{equation}
	where $D$ is the spatial dimension and ${\mathscr N}\left(\epsilon\right)$ is number of available states whose energy is lower or equal to $\epsilon$. Given the nature of the 2-D magnetic lattices in \NaYbLu{}, the spatial dimension can be either 1 or 2. The internal energy of the itinerant entropy carriers can be evaluated as:
	\begin{equation}
		U_{\text{mag, I}}\left(T\right)=\int_{\epsilon=\Delta}^\infty  \frac{\epsilon{\mathscr D}\left(\epsilon\right) d\epsilon}{\exp\left[\epsilon\left/\left(k_BT\right)\right.\right]-1}\sim\int_{\epsilon=\Delta}^\infty  \frac{\epsilon^D d\epsilon}{\exp\left[\epsilon\left/\left(k_BT\right)\right.\right]-1} \text{.}
	\end{equation}
	Here we suggest that the entropy carries follows Bose-Einstein statistics; the subscript ``I'' stand for ``itinerant''. We introduce the energy gap $\Delta$ by setting the lower bound of the integral for energy. Taking a derivative with respect to temperature, we obtain the heat capacity as:
	\begin{equation}
		\begin{split}
			C_{\text{mag, I}}\left(T\right)&=\frac{\partial U_{\text{mag, I}}}{\partial T}\sim\int_{\epsilon=\Delta}^\infty \frac{\epsilon^{D+1}}{k_BT^2}\exp(\frac{\epsilon}{k_BT})\frac{d\epsilon}{\left(\exp\left[\epsilon\left/\left(k_BT\right)\right.\right]-1\right)^2} \\
			&\sim T^D\int_{x=\Delta\left/\left(k_BT\right)\right.}^\infty \frac{x^{D+1}\exp(x)}{\left[\exp(x)-1\right]^2}dx \text{.}
		\end{split}
	\end{equation}
	Applying the kinetic formula, the contribution to the thermal conductivity from itinerant magnetic carriers can be modeled as:
	\begin{equation}\label{Eq-DimensionComparison}
		\begin{split}
			\kappa_{\text{mag}}\left(T\right)&=\frac{1}{3}C_{\text{mag, I}}\left(T\right)l_{\text{mag}}v_{\text{mag}} \\
			&=
			\begin{cases}
				\displaystyle{A_{\text{mag}}T\int_{x=\Delta\left/\left(k_BT\right)\right.}^\infty \frac{x^2\exp(x)}{\left[\exp(x)-1\right]^2}dx} \text{,} & \text{1-dimensional;} \\[15 pt]
				\displaystyle{A_{\text{mag}}T^2\int_{x=\Delta\left/\left(k_BT\right)\right.}^\infty \frac{x^3\exp(x)}{\left[\exp(x)-1\right]^2}dx} \text{,} & \text{2-dimensional.}
			\end{cases}
		\end{split}
	\end{equation}
	Here we assume neither the mean free path nor the group velocity of entropy carrier have temperature or frequency dependency---given the relative narrow range of temperatures where we see the carrier, this assumption is rather reasonable. The 1-dimensional case---as expected---has the same form as the gapped entropy carriers in spin ladders \cite{Sologubenko2000, Hess2001}.
	
	\begin{figure}[t!]
		\begin{center}
			\includegraphics[width = 8.8 cm]{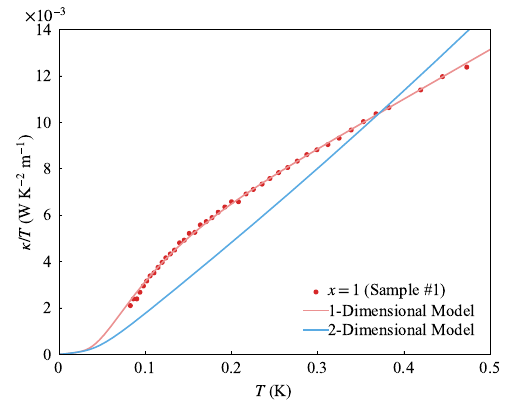}
			\caption{%
				The thermal conductivity of \NaYb{}, along with the 1-D and 2-D modelling of a gapped itinerant magnetic carrier described in Eq.\eqref{Eq-DimensionComparison}.
			}
			\label{Fig-DimensionComparison}
		\end{center}
		\DRule
	\end{figure}
	
	Plotted in \FigRef{Fig-DimensionComparison} is the thermal conductivity of \NaYb{}, as well as the 1-dimensional (1-D) and 2-dimensional models with the addition of a temperature-power-law phonon thermal conductivity. While the 1-D model fits the experimental data well, the signature bump shape cannot be generated---in this linear-linear plot scale---from the 2-D model regardless the values of free fitting parameters. Needless to say, this demonstrates beyond doubt that the gapped itinerant excitation must travel in 1-D space.
	
	In conclusion, the thermal conductivity of \NaYbLu{}---as the combination of both phonon and magnetic parts---can be modelled as:
	\begin{equation}
		\kappa=\kappa_{\mathrm{ph}}+\kappa_{\mathrm{ph}}=A_{\mathrm{ph}}T^\alpha+\displaystyle{A_{\text{mag}}T\int_{x=\Delta\left/\left(k_BT\right)\right.}^\infty \frac{x^2\exp(x)}{\left[\exp(x)-1\right]^2}dx} \text{.}
	\end{equation}
	In the temperature range of our measurements, this 1-D model behaves like the simple exponential expression as in Eq.\eqref{Eq-ThermalConductivity}---as suggested by reference \cite{Yamashita2009}. Smaller fitting errors are observed, however, when this 1-D model is used to fit the data.
	
	\subsection{Discussion on the Origin of the Gap}
	Next, let us discuss in more detail the nature of the gap. As illustrated in the table of \FigRef{Fig-TemperatureEvolution}, we propose two types of low-energy excitations in a VBG that might act as entropy carriers. In the first case, a series of dimers flip along an existing domain wall like dominoes, initiated/terminated by the merging/splitting of a dimer and an orphan spin \cite{Kawamura2019, Kimchi2018}. One can interpret this type of excitation as an orphan spin tunnelling through the tiling of dimers. This is akin to a spinon motion where a pair of end single-site excitations is lined via a tunnel through an all-to-all entangled ocean of resonating valence bonds \cite{Kawamura2019, Savary2016}. In the case of VBG, however, the entanglement only happens locally within each dimer.
	
	We can deconstruct this domino chain reaction into its most basic form---a single dimer flip. Imagine a local ground state where an orphan spin (spin \#1) and a dimer (spins \#2 and \#3) form neighbours. When excited, the dimer splits into 2 orphan spins, one of which (say \#2) recombines with the original orphan spin to form a new dimer (\#1 and \#2). Simply put, an orphan spin plus dimer turns into a dimer plus orphan spin.
	
	Let us take the vacuum state as three free non-interacting spins.  For simplicity, let us limit our Hamiltonian to just NN AFM Heisenberg interactions, and assign interaction strengths such that it is $\mathscr{J}'$ between spin \#1 and \#2, but $\mathscr{J}$ between spin \#2 and \#3 ($\mathscr{J} > \mathscr{J}'$). In the ground state, the dimer formed between spin \#2 and \#3 must be a singlet and from a semi-classical point of view, no magnetic field is generated at the site of the orphan spin \#1, such that its spin levels remain degenerate. Compared to the vacuum state, the ground state energy is thus approximately $-3\mathscr{J}/4$---lowered by the singlet spin configuration within dimer. The energy contribution from the orphan spin should be marginal as it neither interacts nor entangles with the dimer. Similarly, the energy of the excited state is approximately $-3\mathscr{J}'/4$. As a result, the energy gap for the excitation is $3\left(\mathscr{J}-\mathscr{J}'\right)/4$---in other words, the gap is the difference of coupling strengths between the neighbouring bonds, which is determined by the quenched randomness in the lattice.
	
	A similar situation can be outlined for the second case, where as the dimer is absorbed into the neighbouring cluster, the system wave function changes and while some bonds in the new cluster lower the total energy by switching to spin-singlet alignments, some old bonds get effectively deactivated---either because their alignments change or their probability amplitudes in the new wave function vanish. The gap is then the summation of new singlet bonds minus the old ones and again should be largely determined by local inhomogeneity in the bond strengths---at the length scale of the clusters.
	
	The gap we observed in the thermal conductivity is thus a lattice average of all the local inhomogeneity among the bond strengths. While probabilistically, one might be able to find some regions in the lattice where bond energy is homogeneous, thus yielding a phenomenologically gapless itinerant excitation, our data indicates that such regions are relatively rare and sparse---this aligns with our proposal of localized disorder surrounding individual Yb\textsuperscript{$3+$} ions.
	
	\begin{figure}[t!]
		\begin{center}
			\includegraphics[width = 12.1 cm]{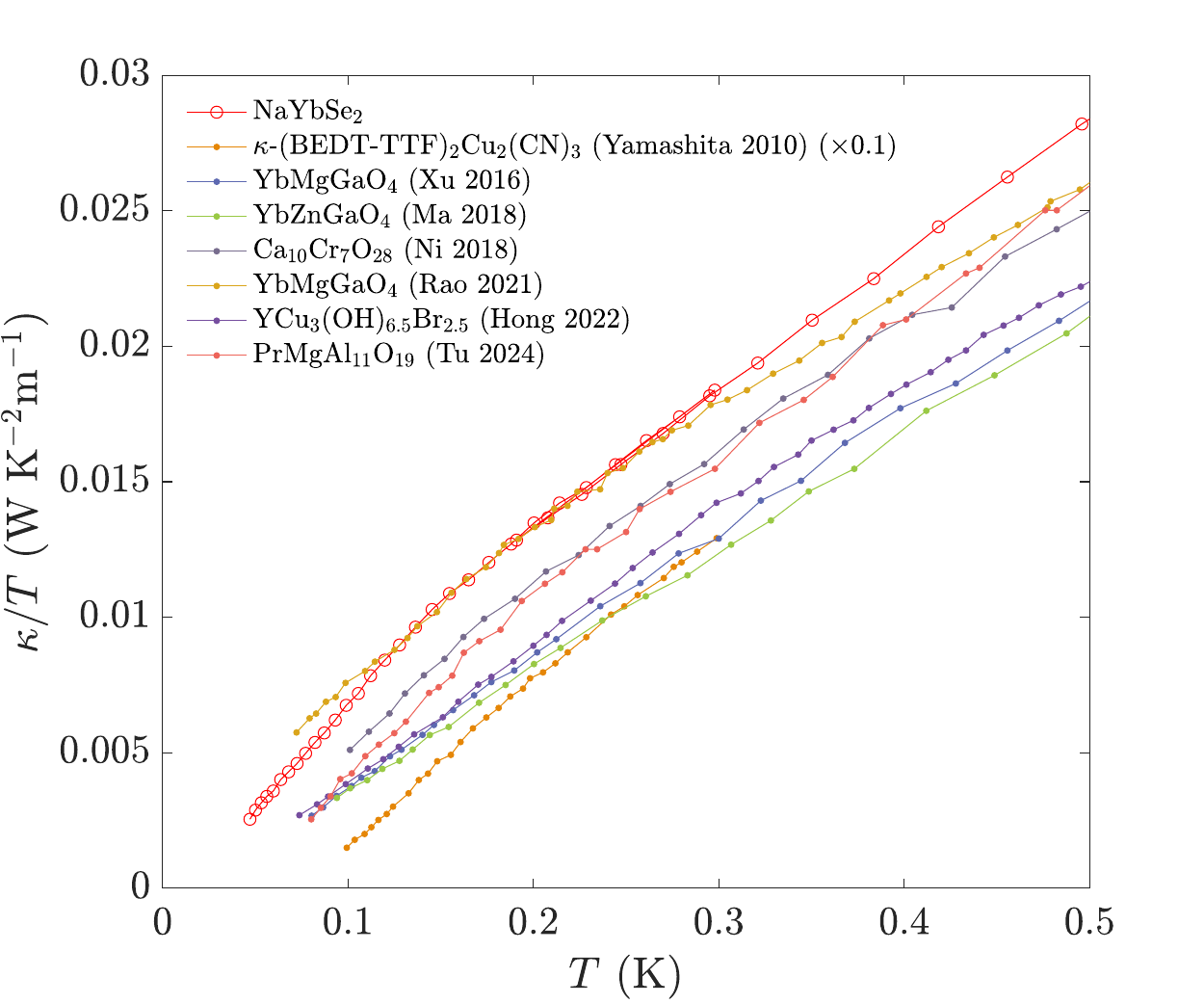}
			\caption{Zero-field thermal conductivity of a variety of related compounds, all of which lack $T$-linear residue term in thermal conductivity but instead show bump-like features. The data is taken from \cite{Yamashita2009, Xu2016, Ma2018, Ni2018, Hong2022, Tu2024}.}
			\label{Fig-Universality}
		\end{center}
		\DRule{}
	\end{figure}
	
	Finally, let us analyse the population and mean free paths of the gapped itinerant excitations. The majority of the entropy release leading to a VBG state comes from the spin freezing/flipping in the formation of dimers and locally entangled clusters. The itinerant excitation described above, however, corresponds to the entropy of all possible tiling configurations of the dimers and clusters---which should have a comparably negligible number of states and consequently a smaller entropy release. This helps to explain the apparent contradiction between heat capacity and thermal transport in a slew of QSL candidates \cite{Yamashita2009, Xu2016, Ma2018, Ni2018, Rao2021, Hong2022, Hong2023, Hong2024, Tu2024, Kimchi2018}---the thermal conductivity of some of which are shown in \FigRef{Fig-Universality}. More specifically, low-energy excitations in large-size clusters may be gapless and popular, and thus generate a temperature power law in heat capacity. Regardless, they are not mobile \cite{Kimchi2018}---limited by the physical size of the corresponding clusters or dimers---and therefore have a miniscule contribution to the thermal conductivity. Only entropy carriers from the reconfiguration of VBG tilings could travel beyond the length scale of individual magnetic features and thus show a gapped behaviour in the thermal conductivity. As a result, instead of decreasing in population as the magnetic lattice get diluted with decreasing $x$ in \NaYbLu{}, the population of itinerant magnetic entropy carriers---$A_{\mathrm{mag}}$ in \FigRef{Fig-ThermalConductivity}~(c)---peaks near the percolation threshold, as there are effectively more boundaries between spin features.
	
	With this in mind, we can estimate a worst-case-scenario lower bound on the mean free path of the itinerant magnetic entropy carriers. Using the kinetic formula, we get:
	\begin{equation}
		l_{\mathrm{mag}}=\frac{3\kappa_{\mathrm{mag}}}{C_\text{mag, I}v_{\mathrm{mag}}}>\frac{3\kappa_{\mathrm{mag}}}{C_{\mathrm{mag}}v_{\mathrm{mag}}} \text{.}
	\end{equation}
	Taking the group velocity $v_{\mathrm{mag}}$ from the non-spin-wave peak in the INS data \cite{Dai2021}, and the measured magnetic heat capacity and thermal conductivity at 200~mK, the mean free path is calculated to be \SI{6.6}{\angstrom}. This is already larger than the nearest-neighbour Yb-separation of $\SI{4.06}{\angstrom}$, and thus demonstrates the feasibility of a magnetic carrier. It should be stressed however, the true mean free path of the itinerant magnetic entropy carrier is expected to be much longer as the heat capacity of itinerant carriers $C_\text{mag, I}$ is much smaller than the overall magnetic heat capacity.
	
	\subsection{Absence of Spin-Phonon Decoupling}
	For an itinerant magnetic entropy carrier to be detected through thermal conductivity measurements, it must induce temperature changes that can be detected by thermometers bonded to the samples. In other words, the spin subsystem hosting the magnetic excitations must couple sufficiently strongly to the phonon subsystem.
	
\begin{figure}[t!]
		\begin{center}
			\includegraphics[width = 8.8 cm]{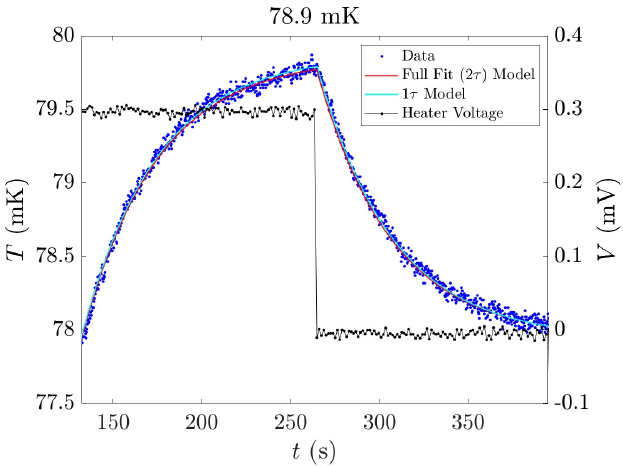}
			\caption{Portion of the relaxation time data for a single data point in the \NaYb{} zero-field heat capacity. The blue points are the sample thermometer temperature, the black points are the heater voltage, and the lines show fits to the single time constant ($1\tau$) and two time constant ($2\tau$) models.}
			\label{Fig-LowTemperatureHeatCapacityPulse}
		\end{center}
		\DRule{}
	\end{figure}

	In previous publications, a spin-phonon decoupling has been put forward as a potential reason for the low-temperature loss of thermal conductivity in QSL candidates \cite{Isono2018, Hong2023}, and also as a general phenomenon in magnetically ordered systems \cite{Sanders1977}. The result of this decoupling might show up as a drop in thermal conductivity at very low temperatures and might visually appears like some gapped exponential behaviour. However, we can discount that interpretation here for two reasons---(i) it cannot explain the bump-like shape at higher $\sim\SI{500}{\milli\kelvin}$ as an addition rather than subtraction on top of a temperature-power-law thermal conductivity, and (ii) the low-temperature specific heat data is well described by a single time constant (\FigRef{Fig-LowTemperatureHeatCapacityPulse}), which implies that the entire system is well coupled even to the lowest measured temperature.
	
	\clearpage
	
	\section{Modelling of Heat Capacity}
	\subsection{Attempts to Constrain the Pseudospin-1/2 Hamiltonian at Low \texorpdfstring{Yb\textsuperscript{$3+$}}{Yb3+} Content}
	
	The heat capacity is defined as the derivative of the internal energy with respect to temperature, and can therefore be calculated once the energy levels of a system are known. In the temperature range of interest the electronic spins cannot be treated as independent and so the number of energy levels will scale as $N^{2S+1}$, where $N$ is the number of interacting magnetic ions and $S$ the spin. Obviously then, the method of exact diagonalisation is completely inappropriate to model a massively entangled, $J=7/2$ system such as \NaYb{}. However, in those compounds with sufficiently low Yb\textsuperscript{$3+$} content, the magnetic clusters are almost entirely small enough that exact diagonalisation is feasible, especially at low temperatures and fields where the pseudospin-1/2 model is a good approximation (see the following section). It should therefore be possible to determine characteristics of the spin Hamiltonian for the entire compositional series by fitting to the thermodynamic properties of the $x\rightarrow 0$ compounds.
	
\begin{figure}[t!]
		\begin{center}
			\includegraphics[width = 18 cm]{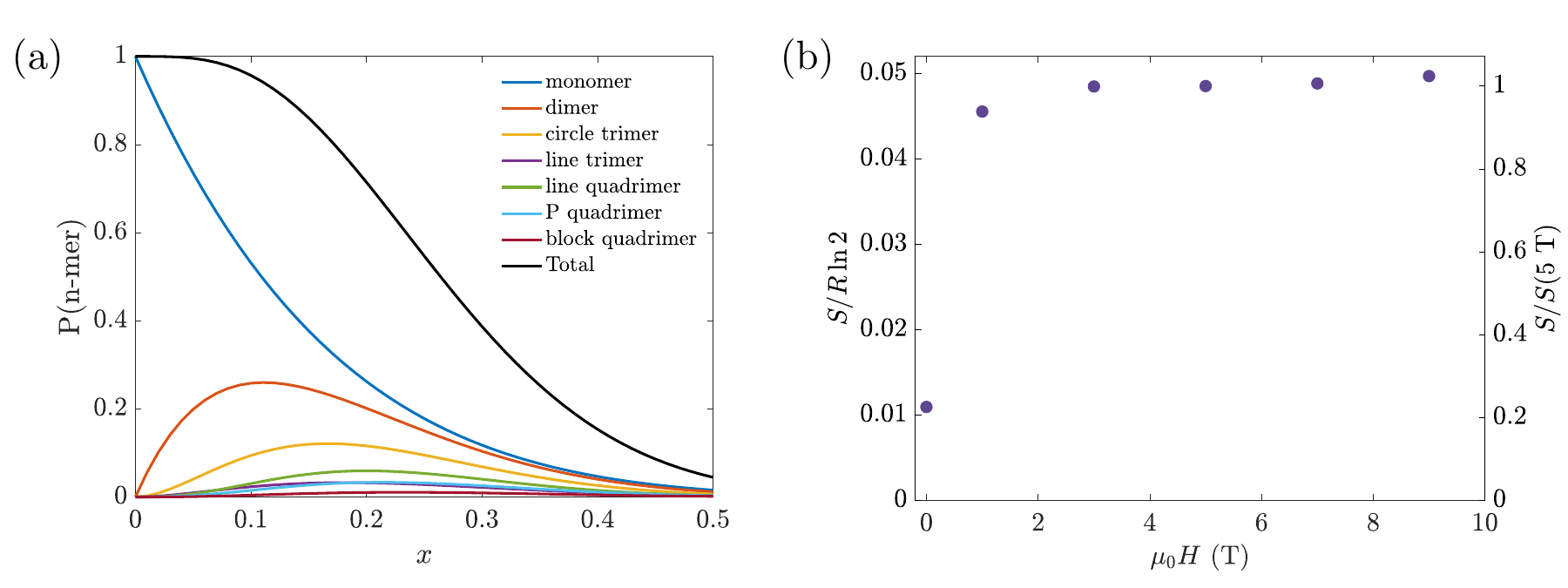}
			\caption{(a) Probability of small magnetic clusters as a function of $x$ (or equivalently proportion of magnetic sites on the triangular lattice). The black line is the sum of all the different n-mers included. (b) Entropy release of NaYb\textsubscript{0.05}Lu\textsubscript{0.95}Se\textsubscript{2} between \SI{500}{\milli\kelvin} and \SI{30}{\kelvin} at a range of fields, as a fraction of $R\ln(2)$ (left axis) and as a fraction of the value (right axis) under the magnetic field of $\mu_0H=\SI{5}{\tesla}$. The corresponding heat capacity data is shown in \FigRef{Fig-Yb5HeatCapacity}.}
			\label{Fig-NMerEntropy}
		\end{center}
		\DRule{}
	\end{figure}

	If we assume perfect randomness of magnetic and non-magnetic sites on the triangular lattice, it is possible to calculate the concentration of each species of magnetic cluster probabilistically. This is shown in \FigRef{Fig-NMerEntropy} (a) for the smallest clusters, and---through the black curve which shows the summed total---illustrates how rapidly the approximation of exclusively small clusters becomes inappropriate as the Yb content increases---this is the reason we limit this analysis to NaYb\textsubscript{0.05}Lu\textsubscript{0.95}Se\textsubscript{2}. It also demonstrates how sensitively the modeling will depend on the Yb content. 
	
\begin{figure}[t!]
		\begin{center}
			\includegraphics[width = 18 cm]{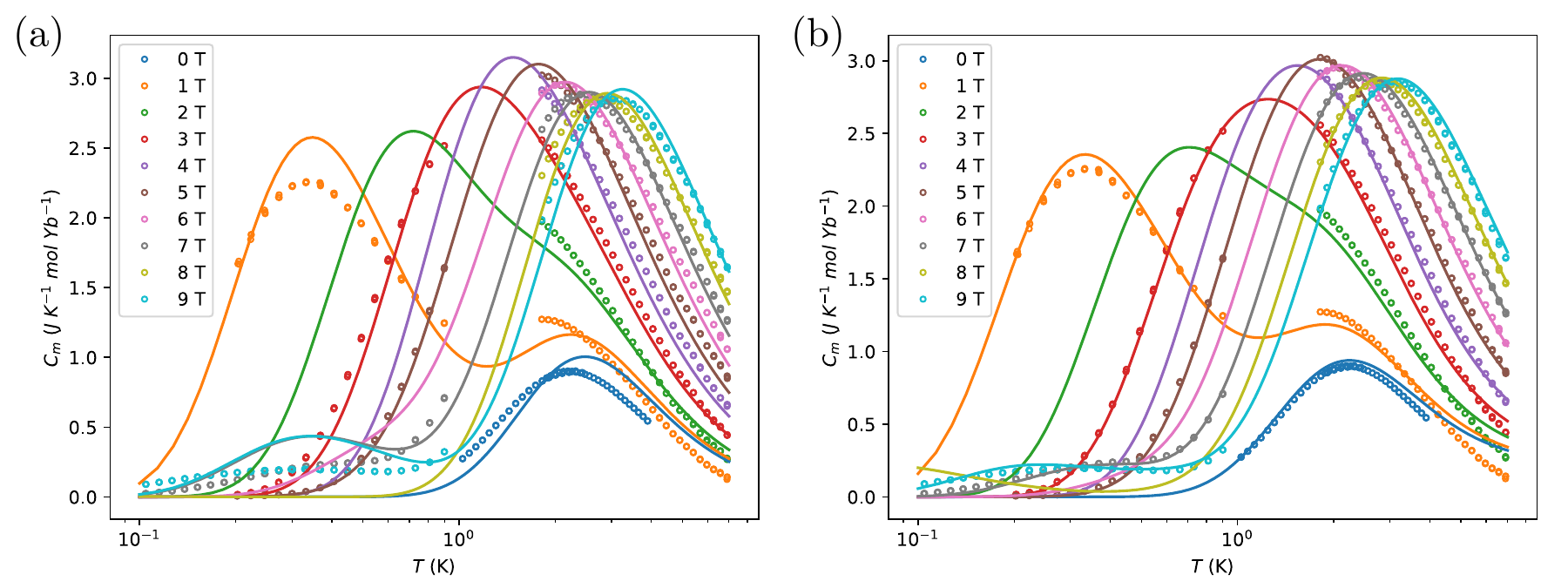}
			\caption{Heat capacity of NaYb\textsubscript{0.05}Lu\textsubscript{0.95}Se\textsubscript{2} for various out-of-plane fields $H_z$. As described in the text, (a) shows fitted curves from a model of isolated ions and a single species of dimer, whereas the curves in (b) are from a model with three species of dimer, with fitted parameters given in Table \ref{Tab-Yb5FitParameters}.}
			\label{Fig-Yb5HeatCapacity}
		\end{center}
		\DRule{}
	\end{figure}

	For the growth intended to be NaYb\textsubscript{0.05}Lu\textsubscript{0.95}Se\textsubscript{2}, energy dispersive X-ray spectroscopy (EDX) measurements give $x=0.05\pm 0.01$. However, we can be more precise by measuring the released entropy, which can be used to calculate the number of spin-1/2 moments. Shown in \FigRef{Fig-NMerEntropy} (b) is the entropy released over the full measured temperature range at a range of fields (taken from the magnetic heat capacity data shown in \FigRef{Fig-Yb5HeatCapacity}). At zero field the isolated ion doublet is degenerate and so any sites without a nearest neighbour will not contribute to the heat capacity, nor the entropy release. However, with the application of field this degeneracy is broken, all sites contribute to the heat capacity, and the released entropy can be used to infer the number of magnetic sites. Importantly, regardless of the proportion of different sized magnetic clusters (and with an exchange strength $\mathscr{J}/k_B\sim\SI{6.1}{\kelvin} $), the full entropy will be released within the measured temperature window at intermediate fields. This analysis gives $x=0.049$ for the measured compound, and therefore (according to the probabilistic calculation) approximately 74\% of magnetic sites will be isolated, 20\% will form nearest neighbour dimers, and the small remainder will form larger magnetic clusters. As a simple check, the measured entropy release in zero-field is approximately 22\% of the intermediate field value, and therefore very close to the expectation. This zero-field value also puts an upper bound on the proportion of larger magnetic clusters, and the potential impact of aggregation of magnetic sites. 
	
	With $x=0.049$ the physical properties can almost entirely be described by a system of isolated ions and nearest-neighbour dimers. In the simplest model, the partition function therefore takes the form:
	\begin{equation}
		Z=Z_1^{(1-x)^6 N}Z_2^{3x(1-x)^8 N} \text{,}
	\end{equation}
	In this equation, $N$ is the number of magnetic ions; partition function
	\begin{equation}
		Z_1 = 2\cosh(\frac{\mu_0\mu_Bg_zH_z}{2k_BT}) \text{,}
	\end{equation}
	is for a single Yb\textsuperscript{$3+$} ion of $g$-factor component $g_z$ under applied out-of-plane magnetic field of strength $H_z$; and
	\begin{equation}
		Z_2=\mathrm{Tr}\left(\exp\left[-\frac{\hat{\mathscr{H}}_2\left(\mathscr{J}_x,\mathscr{J}_z,H_z\right)}{k_BT}\right]\right) \text{;}
	\end{equation}
	is the partition function of a dimer system with Hamiltonian:
	\begin{equation}\label{Eq-DimerHamiltonian}
		\hat{\mathscr{H}}_2(\mathscr{J}_x,\mathscr{J}_z,H_z) = \mathscr{J}_x (\hat{S}_{1,x}\hat{S}_{2,x}+\hat{S}_{1,y}\hat{S}_{2,y})+\mathscr{J}_z\hat{S}_{1,z}\hat{S}_{2,z}+\mu_0\mu_B g_z\left(\hat{S}_{1,z}+\hat{S}_{2,z}\right)H_z \text{;}
	\end{equation}
	where $\hat{S}_{i,\alpha}$ is the $\alpha$-direction component of the pseudospin-$1/2$ operator at site $i$. The best fit is obtained as $\mathscr{J}_x=\SI{0.68}{\milli\electronvolt}$, $\mathscr{J}_z=\SI{0.47}{\milli\electronvolt}$ and $g_z=1.24$. As shown in \FigRef{Fig-Yb5HeatCapacity} (a), this captures the qualitative features of the heat capacity but fails in a quantitative manner.
	
	This qualitative success of this simple model motivates us to consider a semi-phenomenological model:
	\begin{equation}\label{Eq-DimerHamiltonianMoreComponents}
		Z = Z_1^{p_1 N} Z_{2a}^{p_{2a} N}Z_{2b}^{p_{2b} N}Z_{2c}^{p_{2c} N},
	\end{equation}
	where $0<p_1$, $p_{2a}$, $p_{2b}$, $p_{2c}<1$ are phenomenological parameters describing the proportion of each configuration: ion $Z_1$, and dimers $Z_{2a}$, $Z_{2b}$ and $Z_{2c}$. We assume that the three types of dimer are described by a Hamiltonian of the form Eq.~\eqref{Eq-DimerHamiltonian}, but with different exchange parameters. We then fit the specific heat corresponding to the model Eq.~\eqref{Eq-DimerHamiltonianMoreComponents} against the population and exchange parameters, and also the $g$-factor component $g_z$. The fitting results are presented in \FigRef{Fig-Yb5HeatCapacity} (b) and the corresponding fitted parameters are given in Table~\ref{Tab-Yb5FitParameters}---clearly in this case the fits are much improved. 
	
	Obviously with so many free parameters there is little use in attempting to project this model onto the full compound, but the relative success points to there being some broadened energy scale for dimer formation. This is presumably related to disorder, either through a small concentration of impurity moments or site mixing of Yb\textsuperscript{$3+$} and Na\textsuperscript{$ +$} \cite{Dai2021}. Both effects will be comparatively more prominent at low Yb\textsuperscript{$3+$} content, but as mentioned, this is a prerequisite for the method of exact diagonalisation to be feasible. Further, we have made no attempt to account for any deviations from the pseudospin-$1/2$ model, the validity of which is discussed in the next section. Regardless, the failure of the simple model here has implications for interpretations of the full magnetic compound, where the signatures of quantum spin liquid physics are subtle and might be impacted by similar issues.
	
	\begin{table}
		\begin{center}
			\caption{Values of the parameters that give the best fit within model as Eq.~\eqref{Eq-DimerHamiltonianMoreComponents}, shown by the curves in \FigRef{Fig-Yb5HeatCapacity} (b). The exchange couplings are given in units of \si{\milli\electronvolt}.}
			\label{Tab-Yb5FitParameters}
			\begin{tabularx}{\textwidth}{c|X|X|X|X|X|X|X|X|X|X|X}
				\hline
				Parameter & $g_z$ & $p_1$ & $p_2$ & $p_{2,d}$ & $p_{2,n}$ & $\mathscr{J}_x$ & $\mathscr{J}_z$ & $\mathscr{J}_{x,d}$ & $\mathscr{J}_{z,d}$ & $\mathscr{J}_{x,n}$ & $\mathscr{J}_{z,n}$ \\
				\hline\hline
				Best fit & $1.11$ & $0.60$ & $0.060$ & $0.033$ & $0.096$ & $0.483$ & $0.58$ & $0.457$ & $1.06$ & $1.91$ & $-2.75$\\
				\hline
			\end{tabularx}
		\end{center}
		\DRule{}
	\end{table}
	
	\subsection{The Nuclear Heat Capacity of \protect\texorpdfstring{\NaYb{}}{NaYbSe2}}
	\begin{figure}
		\begin{center}
			\includegraphics[width = 8.8 cm]{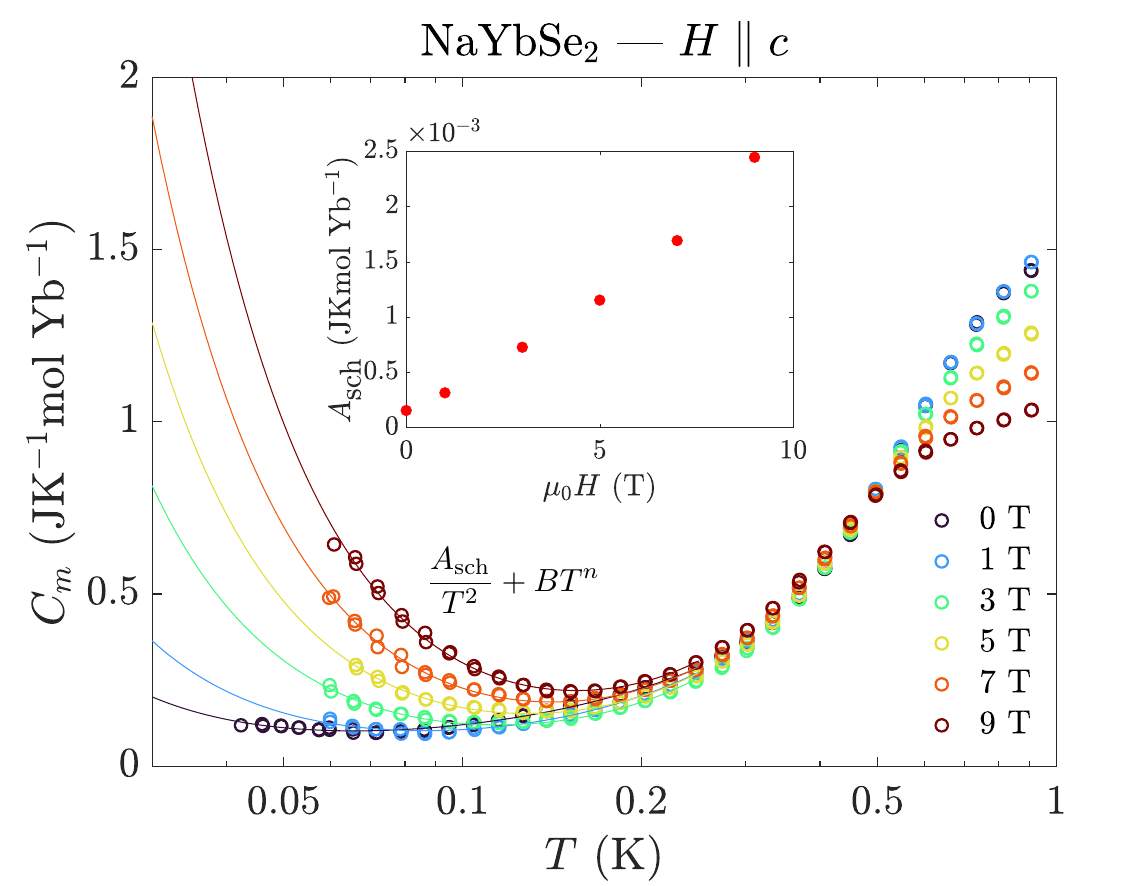}
			\caption{Low-temperature heat capacity of \NaYb{} at a range of fields, applied parallel to $c$-axis. The curves show fits to $A_\text{sch}T^{-2} + BT^n$, and the inset shows the evolution of $A_\text{sch}$ with applied field.}
			\label{Fig-NaYbSe2Schottky}
		\end{center}
		\DRule{}
	\end{figure}
	
	\FigRef{Fig-NaYbSe2Schottky} shows the low-temperature heat capacity of \NaYb{} at a range of fields. This is dominated by an upturn at the lowest temperatures, the magnitude of which grows with increasing field. This upturn can be attributed to the nuclear contribution, with the nuclear spin energy levels being split either through dipolar coupling to the applied magnetic field, quadrupolar coupling to the surrounding crystal field, or either coupling to any unfilled orbitals. As the splitting is $\sim\SI{1}{\milli\kelvin}$---far below the measured temperature range---we only capture the high-temperature tail. This will follow a $T^{-2}$ behaviour regardless of the precise nature of the nuclear subsystem, and as such we fit the low-temperature tail to $A_\text{sch}T^{-2} + BT^n$, where the latter term is some generic power law to account for the unknown behaviour of the underlying magnetic heat capacity. These fits are included in \FigRef{Fig-NaYbSe2Schottky}, and the extracted Schottky coefficient $A_\text{sch}$ is plotted as a function of applied field in the inset. As anticipated from a Zeeman splitting of the nuclear energy levels, $A_\text{sch}$ rises approximately linearly with field, which gives some confidence in the fitting procedure. In zero-field, the nuclear heat capacity is non-zero and $A_\text{Sch}=\SI{1.55e-4}{\joule\kelvin^{-4}\mole^{-1}}$. If we approximate this contribution as arising due to a two-level system with an energy separation of $\delta$, then $A_\text{Sch}=R\delta^2/4$ and $\delta/k_B\sim\SI{10}{\milli\kelvin} $.
	
\begin{figure}[t!]
		\begin{center}
			\includegraphics[width = 18 cm]{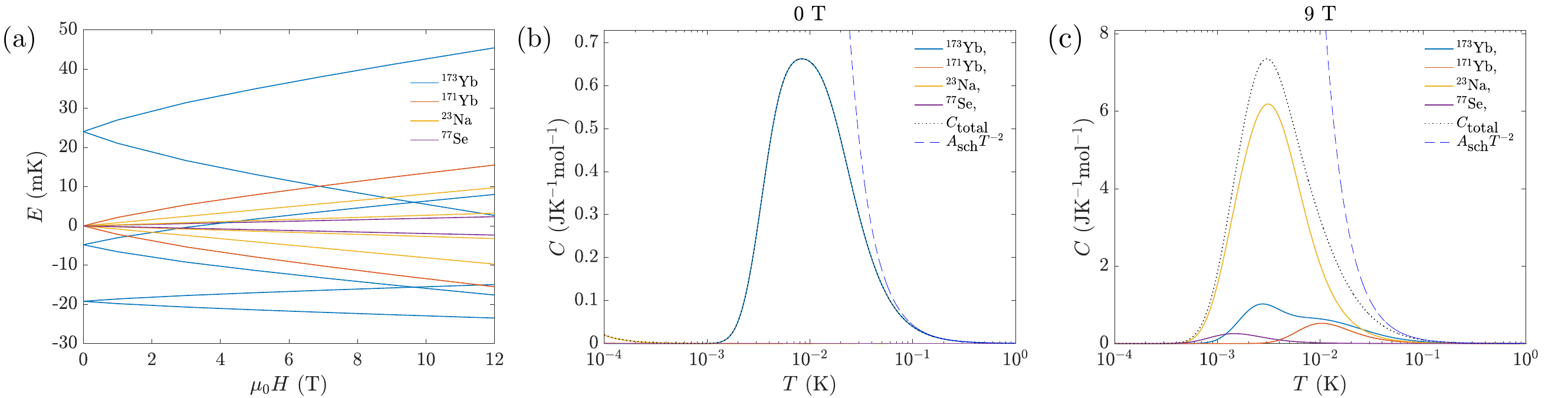}
			\caption{(a) Evolution of the nuclear spin energy levels of all of the isotopes in \NaYb{} as a function of field, according to Eq.~\eqref{Eq-NuclearEnergySplitting} and using the physical parameters described in the text and in Table~\ref{Table-NuclearParameters}. For the Yb site the calculations were performed using the crystal field parameters from \cite{Zhang2021-CEF}. (b) The resultant heat capacity of each isotope at zero-field, also included is the total heat capacity and a $A_\text{sch}T^{-2}$ fit to the high-temperature tail. (c) The same plot at $\mu_0H=\SI{9}{\tesla}$.}
			\label{Fig-NuclearSchottkyInField}
		\end{center}
		\DRule
	\end{figure}

	In order to extract more information from the magnitude of the nuclear contribution---particularly in zero field---we must consider the origin of the splitting. For fields oriented parallel to the symmetry axis, a nuclear spin $I$ will have energy levels
	\begin{equation}\label{Eq-NuclearEnergySplitting}
		E_m=\frac{e^2qQ}{4I\left(2I-1\right)}\left[3m_I^2-I\left(I+1\right)\right]-\gamma_n\hbar H_\text{eff} m_I \text{.}
	\end{equation}
	In this expression, $m_I$ is the azimuthal quantum number, $Q$ the quadrupole moment of the nuclear spin, $eq$ the electric field gradient (EFG) at the nucleus (or $V_{zz}$), $\gamma_n$ the gyromagnetic factor and $H_\text{eff}$ the effective field at the nucleus. All of the nuclear isotopes contained in \NaYb{} are listed in Table~\ref{Table-NuclearParameters}, but by far the largest zero-field contribution will be from the $^{173}$Yb, due to interactions between the nuclear spin and the unfilled $4f$ orbital (see \FigRef{Fig-NuclearSchottkyInField} for simulated data across the full measured field range). In this case, $H_\text{eff}=H+A\Braket{\hat{J}_z}$, where $H$ is the applied field, $A$ the hyperfine coupling and $\Braket{\hat{J}_z}$ the expectation value of the $4f$ orbital. Therefore, in order to calculate the energetic configuration of the nuclear levels---and corresponding nuclear heat capacity---we must determine the EFG at the Yb site and the effective moment of the unfilled $4f$ orbital ($\mu_\text{eff}=g_J\Braket{\hat{J}_z}$). The other unknowns---$Q$, $\gamma_n$ and $A$---are isotope dependent and well characterised (see Table \ref{Table-NuclearParameters}).
	
	Typically the EFG for a chosen ion at a specific crystallographic site is measured via NMR or M\"{o}ssbauer spectroscopy, but unfortunately there are no such studies for the Yb ion in \NaYb{} (NMR has been performed but this focused on $^{23}$Na \cite{Ranjith2019, Zhu2023}). The EFG must instead be estimated from the crystal field parameters and derived wavefunctions \cite{Stewart1985}, using
	\begin{equation}
		V_{zz}=-\frac{4B^0_2(1-\gamma_\infty)}{\abs{e}\alpha_J\Braket{r^2}_{4f}(1-\sigma_2)}-\frac{\alpha_J\abs{e}\left(1-R_Q\right)}{4\pi\epsilon_0}\Braket{\frac{1}{r^3}}_{4f}\Braket{\hat{O}^0_2}.
	\end{equation}
	In this expression, $\sigma_2$, $\gamma_\infty$ and $R_Q$ parametrise the Sternheimer shielding, antishielding and atomic shielding respectively, $\langle r^n \rangle_{4f}$ are the expectation values of the $4f$ radial wavefunction, and $\alpha_J$ is the Stevens factor, all of which have been calculated previously for Yb\textsuperscript{$3+$} \cite{Elliott1953,Freeman1962,Gupta1973}. As discussed above, $B^0_2$ is a crystal field parameter and $\Braket{\hat{O}^0_2}$ is the expectation value of the Stevens operator \cite{Hutchings1964}, such that the first term describes the contribution to the EFG from the lattice and second the contribution from the unfilled $4f$ orbital. 
	
	As a first approximation---and upper bound---we can assume that (i) the latter term will dominate (this is generally true for Yb compounds) and (ii) the magnetisation is saturated (such that $\Braket{\hat{O}^0_2}=J\left(2J-1\right)=21$). This is the case treated in \cite{Bleaney1963} and results in an EFG of \SI{-8.1e22}{\volt\per\meter\squared}, or splitting between the $m_I=\pm5/2$ and $m_I=\pm1/2$ levels of \SI{118}{\milli\kelvin}.  Clearly, and as anticipated, this method gives a huge overestimation of the EFG---recall, the measured zero-field splitting assuming a two-level system is just \SI{10}{\milli\kelvin}. As a more realistic second attempt, we can use the previously estimated crystal field parameters for \NaYb{} \cite{Zhang2021-CEF}. In this case the electric field gradient is calculated to be \SI{-2.96e22}{\volt\per\meter\squared}, which corresponds to a total splitting of 43~mK. Again, this is a fairly large overestimation and suggests some error in those crystal field parameters. In order to match the observed $A_\text{Sch}=\SI{1.55e-4}{\joule\kelvin^{-4}\mole^{-1}} $, the EFG at the Yb site should be \SI{-2e22}{\volt\per\meter\squared}. Note however, the expectation value $\Braket{\hat{O}^0_2}$ requires knowledge of the precise wavefunctions, which will depend on all six crystal field parameters. This estimation of the EFG therefore cannot effectively constrain their values, in isolation at least.
	
\begin{figure}[t!]
		\begin{center}
			\includegraphics[width = 18 cm]{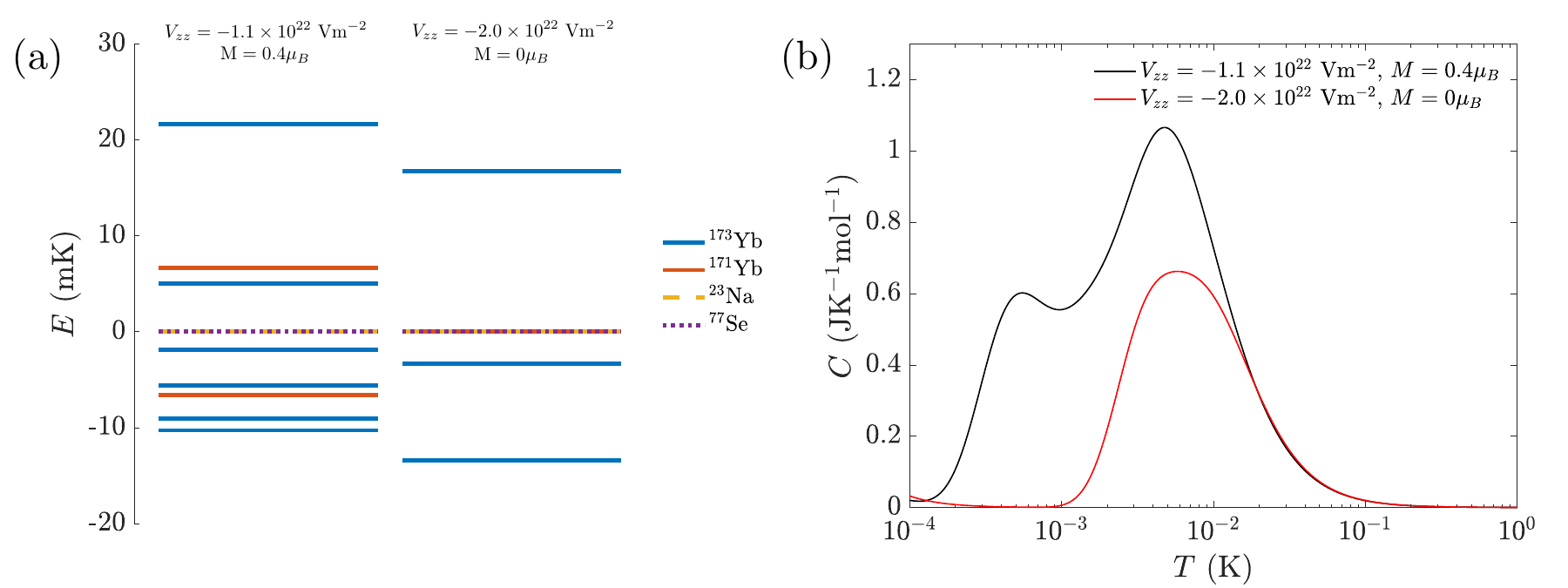}
			\caption{(a) Zero-field nuclear spin energy levels for the isotopes in \NaYb{}, according to the values in Table \ref{Table-NuclearParameters} and for two different strengths of electric field gradient and Yb static moment, both of which describe the measured data equally well. (b) The resultant heat capacity in the two scenarios, where no discernable difference between the two can be seen in the temperature range where our measurements take place.}
			\label{Fig-NuclearSchottkyStaticMoment}
		\end{center}
		\DRule{}
	\end{figure}

	Finally, as mentioned, the nuclear spin energy levels will also be Zeeman split by the effective field $H_\text{eff}=H+A\Braket{\hat{J}_z}$, which is a sum of the applied field and effective moment of the Yb $4f$ orbital. This implies that the above EFG will be an overestimation in the presence of any static moment. Further, \FigRef{Fig-NuclearSchottkyStaticMoment} demonstrates that although a Zeeman splitting will qualitatively change the behaviour of the nuclear heat capacity, the high-temperature tail will appear similar regardless of the magnitude of the static moment (with appropriately scaled EFG to match the data). Therefore, once again, in order to make any claims with regards to the zero-field static moment a more accurate determination of the crystal field parameters is necessary. The best we can say is that, if there is no static moment for \NaYb{} in zero field---as the data would suggest---the EFG at the Yb\textsuperscript{$3+$} site will be $\sim\SI{-2e22}{\volt\per\meter\squared}$.
	
	\begin{table}
		\begin{center}
			\caption{Nuclear properties of the relevant isotopes in \NaYbLu{}. References given in the column headers, aside from the electric field gradient which has been calculated for Yb and Lu using the crystal field parameters from \cite{Zhang2021-CEF}.}
			\label{Table-NuclearParameters}
			
			\begin{tabularx}{\textwidth}{c|c|Y|Y|Y|Y|Y}
				\hline
				Isotope & $I$ & Natural Abundance \newline (\%) \cite{deLaeter2003-Isotopes} & $\gamma$ \newline (\si{\mega\hertz\per\tesla}) \cite{Stone2014} & $A$ \newline (\si{\tesla\mu_B^{-1}})  \newline \cite{Bleaney1963} & $Q$ \newline ($\times10^{-28}$ m$^2$)  \newline \cite{Stone2014} & $V_{zz}$  \newline (\si{\volt\per\meter\squared}) \\ \hline \hline
				\textsuperscript{171}Yb & $1/2$ & $14.3$ & $7.5052$ & $118$ & - & $\SI{-2.8e22}{}$ \\
				\hline
				\textsuperscript{173}Yb & $5/2$ & $16.1$& $-2.0672$ & $118$ & $2.80$ & $\SI{-2.8e22}{}$ \\
				\hline
				\textsuperscript{23}Na & $3/2$ & $100$ & $11.26873$& - & $0.104$ & \SI{1.6e20}{} \cite{Ranjith2019} \\
				\hline
				\textsuperscript{77}Se & $1/2$ & $7$ & $8.13422$ & - & - & - \\
				\hline
				\textsuperscript{175}Lu & $7/2$ & $97.4$ & $4.8473$ & - & $3.49$ & \SI{2.2e21}{} \\
				\hline
				\textsuperscript{176}Lu & $7$ & $2.6$ & $3.4411$ & - & $4.92$ &\SI{2.2e21}{} \\
				\hline
			\end{tabularx}
		\end{center}
		\DRule{}
	\end{table}
	
	\clearpage
	
	\section{Conditions for Failure of the Pseudospin-1/2 Model}
	In \NaYbLu{}, the lowest order splitting of the $4f^{13}$ energy levels of the Yb\textsuperscript{$3+$} is due to the surrounding crystal field. The Yb\textsuperscript{$3+$} site has $C_\text{3v}$-symmetry, such that the crystal field Hamiltonian is Eq.\eqref{Eq-CEFTrigonal}. Symmetry dictates that the lowest lying $J=7/2$ state will be split into four time-reversed doublets, with one doublet being the pure $m_j=3/2$ state, and other the three doublets being superpositions of the $m_j=1/2,~5/2,~7/2$ states. The crystal field parameters will determine the degree of splitting and admixing, and by extension the thermodynamic properties of the magnetic subsystem. Zhang et al. \cite{Zhang2021-CEF} have estimated the crystal field parameters in \NaYb{} via inelastic neutron scattering and Raman scattering, and found that the lowest energy doublet is separated from the first excited state by $\sim180$~K. This implies that at sufficiently low temperatures and fields, the magnetic behaviour of \NaYbLu{} is determined almost entirely by the properties of this lowest energy Kramer's doublet, which can be treated as a pseudospin-1/2 moment.

\begin{figure}[t!]
		\begin{center}
			\includegraphics[width = 8.8 cm]{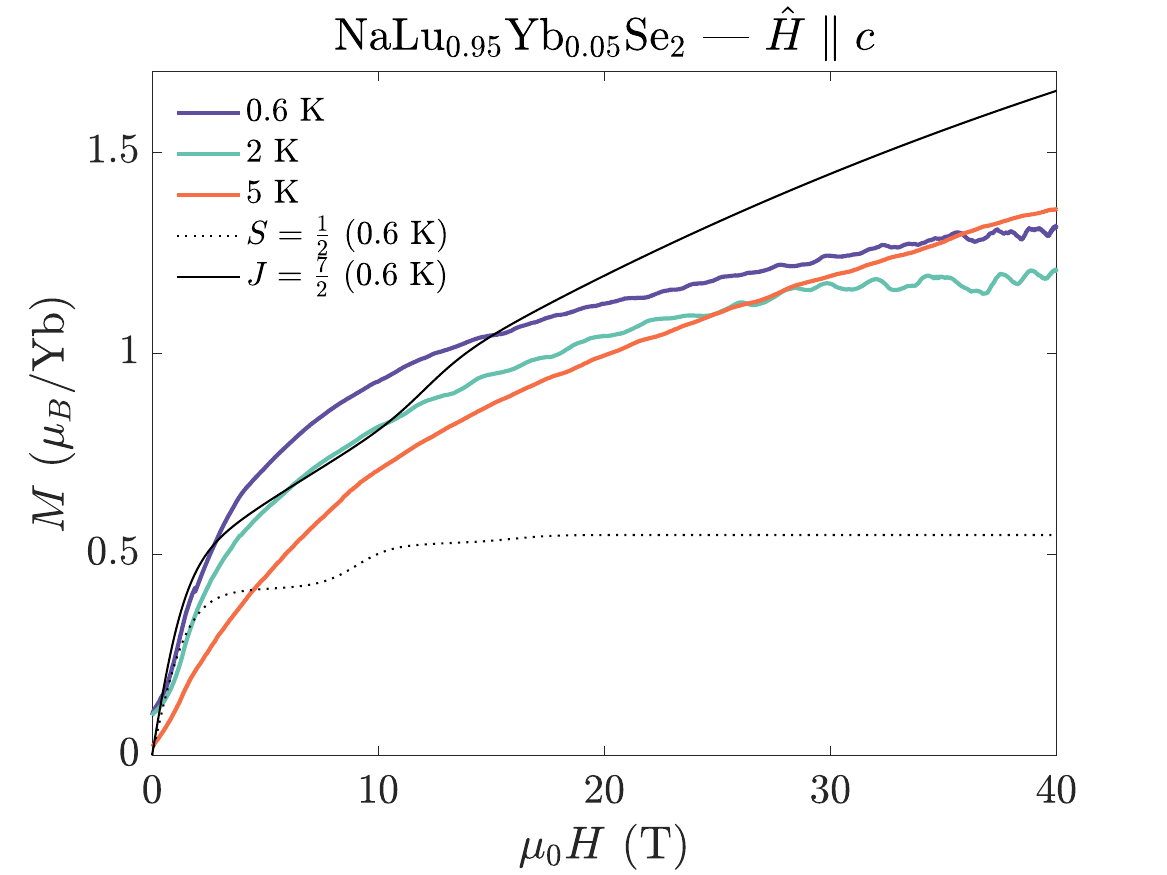}
			\caption{High-field magnetisation of NaYb\textsubscript{0.05}Lu\textsubscript{0.95}Se\textsubscript{2} for out-of-plane fields. The dotted line shows the magnetisation calculated for the pseudospin-$1/2$ model at \SI{600}{\milli\kelvin}, with the proportion of different sized clusters and parameters as described in the text for NaYb\textsubscript{0.05}Lu\textsubscript{0.95}Se\textsubscript{2}. The solid line shows the calculated magnetisation for the full $J=7/2$ model, also at \SI{600}{\milli\kelvin}, using the coefficients estimated from INS data \cite{Zhang2021-CEF}.}
			\label{Fig-Yb5Magnetization}
		\end{center}
		\DRule{}
	\end{figure}

	\FigRef{Fig-Yb5Magnetization} shows magnetisation data for NaYb\textsubscript{0.05}Lu\textsubscript{0.95}Se\textsubscript{2}, for $\mathbf{H}\parallel c$ and at temperatures of \SI{600}{\milli\kelvin}, \SI{2}{\kelvin}, and \SI{5}{\kelvin}. Also included is a plot of the calculated pseudospin-$1/2$ magnetisation at \SI{600}{\milli\kelvin}, assuming the proportion of magnetic clusters from the probabilistic calculation shown in \FigRef{Fig-NMerEntropy} (a), and the calculated $g$-factor and exchange strengths from the simplified model (Eq.~\eqref{Eq-DimerHamiltonian}). It is clear that this pseudospin-1/2 treatment becomes rapidly inadequate, saturating above $\sim\SI{10}{\tesla}$ while the measured data continues to rise in a monotonic fashion, far beyond the saturation value. This demonstrates that while the pseudospin-1/2 model does well in describing the properties of \NaYb{}, the influence of the full spin cannot be discounted even at reasonably low temperatures and fields. Again, this should be relevant in trying to interpret subtle experimental features in the full magnetic compound.

	The continuous increase in the magnetisation is readily explained through excitations to higher crystal field levels. This is typically accounted for by the addition of a small linear (Van Vleck) contribution---and indeed this is valid for $\mathbf{H} \perp c$ \cite{PritchardCairns2022}---but evidently that is not appropriate here. Instead, included in \FigRef{Fig-Yb5Magnetization} is the calculated magnetisation for the full $J=7/2$ spin, using the estimated crystal field parameters from Zhang et al. \cite{Zhang2021-CEF}. This qualitatively matches the experimental data but struggles to reproduce the magnitude, nor the high-field curvature. The absence of the intermediate field kink in the measured data---which corresponds to the dimer triplet energy level moving below the singlet---is also puzzling. We expect the crystal field at the Yb site to be similar regardless of composition, as the unit cell volume changes by $<0.7$\% between \NaYb{} and \NaLu{} \cite{PritchardCairns2022}. Also, at this composition the magnetisation will be dominated by single ion physics, such that errors in estimating the exchange Hamiltonian should not have a significant impact. We therefore ascribe the discrepancy between the measured data and calculated full spin magnetisation to errors in the crystal field parameters. This conclusion is given further support in the following section, where we demonstrate that this choice of crystal field parameters wildly overestimates the nuclear contribution to the heat capacity. An accurate determination of the crystal field parameters in rare-earth systems is notoriously difficult, especially in this case which requires fitting six independent variables. However, it is evident from the magnetisation data that any attempts to model \NaYbLu{} will require the full $J=7/2$ spin even in the presence of small fields, and accurate crystal field parameters are thus essential.
	
	\clearpage
	
	\beginreferences

\end{document}